*Article*

# Elastic Information Bottleneck

**Yuyan Ni [1], Yanyan Lan [2],\*, Ao Liu [3] and Zhiming Ma [1]**

[1] Academy of Mathematics and Systems Science, Chinese Academy of Sciences, Beijing 100190, China
[2] Institute for AI Industry Research, Tsinghua University, Beijing 100084, China
[3] School of Computer Science and Technology, University of Chinese Academy of Sciences, Beijing 100049, China
\* Correspondence: lanyanyan@tsinghua.edu.cn

**Abstract:** Information bottleneck is an information-theoretic principle of representation learning that aims to learn a maximally compressed representation that preserves as much information about labels as possible. Under this principle, two different methods have been proposed, i.e., information bottleneck (IB) and deterministic information bottleneck (DIB), and have gained significant progress in explaining the representation mechanisms of deep learning algorithms. However, these theoretical and empirical successes are only valid with the assumption that training and test data are drawn from the same distribution, which is clearly not satisfied in many real-world applications. In this paper, we study their generalization abilities within a transfer learning scenario, where the target error could be decomposed into three components, i.e., source empirical error, source generalization gap (SG), and representation discrepancy (RD). Comparing IB and DIB on these terms, we prove that DIB's SG bound is tighter than IB's while DIB's RD is larger than IB's. Therefore, it is difficult to tell which one is better. To balance the trade-off between SG and the RD, we propose an elastic information bottleneck (EIB) to interpolate between the IB and DIB regularizers, which guarantees a Pareto frontier within the IB framework. Additionally, simulations and real data experiments show that EIB has the ability to achieve better domain adaptation results than IB and DIB, which validates the correctness of our theories.

## 1. Introduction

Representation learning has recently become a core problem in machine learning, especially with the development of deep learning methods. Different from other statistical representation learning approaches, the information bottleneck principle formalizes the extraction of relevant features about $Y$ from $X$ as an information-theoretic optimization problem: $\min_{p(t|x)} L = \min_{p(t|x)} f(X;T) - \beta g(Y;T)$, where $p(t|x)$ amounts to the encoder of the input signal $X$, $f$ stands for the compression of representation $T$ with respect to input $X$, $g$ stands for the preserved information of $T$ with respect to output $Y$, $Y \leftrightarrow X \leftrightarrow T$ forms a Markov chain, and $\beta$ is the tradeoff parameter. The basic idea of the information bottleneck principle is to obtain the information that X provides about Y through a 'bottleneck' representation T. The Markov constraint requires that T is a (possibly stochastic) function of X and can only obtain information about Y through X.

Under this principle, various methods have been proposed, such as information bottleneck (IB) [1], conditional entropy bottleneck (CEB) [2], Gaussian IB [3], multivariate IB [4], distributed IB [5], squared IB [6], deterministic information bottleneck (DIB) [7], etc. Almost all previous methods use the mutual information $I(Y;T)$ as the information preserving function $g$. As for the compression function $f$, there are two typical functions, which categorize these methods into two groups. The first group uses the mutual information $I(X;T)$, a common measure of representation cost in channel coding, as the compression function. Typical examples include IB, CEB, Gaussian IB, multivariate IB, squared IB, etc. Please note that CEB uses the conditional mutual information $I(X;T|Y)$ as the compression function, however, it has been proven to be equivalent to mutual information. Similarly, squared IB uses the square of the mutual information $I^2(X;T)$ as the compression function; still, we put it into the same category. Instead of the mutual information, the second group, including DIB, uses entropy $H(T)$ as the compression function, which is another common measure to quantify the representation cost as in source coding. The reason for this is that entropy is directly related to the quantity to be constrained such as the number of clusters, which is expected to achieve better compression results.

IB has been extensively studied over recent years in theory and application. In theory, IB has been proven to be able to enhance generalization [8] and adversarial robustness [9] by providing a generalization bound and an adversarial robustness bound. In application, IB has been successfully applied in evaluating the representations learned by deep neural networks (DNN) [10–12] and completing various tasks such as geometric clustering by iteratively solving a set of self-consistent equations to obtain the optimal solution of IB optimization problem [13,14], classification [2,15], and generation [16] by serving as the loss function of DNN via variational methods. Recent research also shows that if we merge the loss function of IB with other models, such as BERT and Graph Neural Network, better generalization and adversarial robustness results will be obtained [9,17]. Moreover, IB inspires new methods that improve generalization [18–20]. Likewise, DIB has been applied in geometric clustering [21] and has the potential to be used in similar applications to IB. More wide-ranging work on IB in deep learning and communication is comprehensively summarized in the surveys found in [22–24].

Still, the theoretical results are only valid when the training and test data are drawn from the same distribution, but this is rarely the case in practice. Therefore, it is unclear, especially theoretically, whether IB and DIB are able to learn a representation from a source domain that performs well on the target domain. Moreover, it is worth studying which objective function is better. This is exactly the motivation of this paper. To this end, we formulate the problem as transfer learning, and the target domain test error could be decomposed into three parts: the source domain training error, the source domain generalization gap (SG), and the representation discrepancy (RD), based on the transfer learning theory [25]. Without loss of generality, we assume that both IB and DIB have the ability to achieve small source domain training errors, so our goals become calculating SG and RD and comparing the two methods on these terms.

For SG, Shamir et al. [8] have provided an upper bound related to $I(X;T)$, indicating that minimizing $I(X;T)$ leads to better generalization. However, this theory is not applicable for comparing IB and DIB because DIB minimizes $H(T) = I(X;T) + H(T|X)$, and it is unclear whether further minimizing $H(T|X)$ may bring some advantages. Therefore, we need to derive a new bound. The difficulty lies in that the new bound needs to not only include both $I(X,T)$ and $H(T|X)$ for convenient comparison but also be tighter than the previous one. Since the previous bound in Shamir et al. [8] was represented as a $\phi$ function of the variance, we tackle this problem by introducing a different analysis of these two factors. Specifically for the variance, different from relating the variance to the $L_1$ distance; then, to KL-divergence; and lastly, to mutual information in the previous proof, we bound the variances by functions of expectations and successfully relate them to entropy $H(T)$. Furthermore, we prove a tighter bound for the $\phi$ function. Consequently, we prove a tighter generalization bound, suggesting that minimizing $H(T)$ is better than $I(X;T)$. Therefore, our results indicate that DIB may generalize better than IB in the source domain.

As for RD, it is measured by the $\mathcal{H}\Delta\mathcal{H}$-distance, as in Ben-David et al. [26]. However, this term is difficult to compute because $\mathcal{H}$ is the hypothesis space of classifiers and is diverse for different models. Inspired by the fact that IB and DIB solutions are mainly different with the variance in the representations, we assume that data are generated with a Gaussian distribution. Therefore, we define a pair-wise $L_1$ distance to bound the $\mathcal{H}\Delta\mathcal{H}$- distance and relate RD with the variance in representations. Specifically, IB's representations have larger randomness and thus have smaller RDs. Moreover, the closer the two domains are, the greater the difference between IB and DIB on RD is.

From the above theoretical findings, we conclude that there exists a better objective function under the IB principle. However, how to obtain the optimal objective function remains a challenging problem. Inspired by the trade-off between SG and RD, we propose an elastic information bottleneck (EIB) to interpolate between IB and DIB, i.e., $\min L_{EIB} = (1 - \alpha)H(T) + \alpha I(X;T) - \beta I(Y;T)$. We can see that EIB includes IB and DIB as special cases. In addition, we provide a variational method to optimize EIB by DNN. We conduct both simulations and real data experiments in this paper. Our results show that EIB is more flexible to different kinds of data and achieves better accuracy than IB and DIB on classification tasks. We also provide an example of combining EIB with a previous domain adaptation algorithm by substituting the cross entropy loss with the EIB objective function, which suggests a promising application of EIB. Our contributions are summarized as follows:

- We derive a transfer learning theory for IB frameworks and find a trade-off between SG and RD. Consequently, We propose a novel representation learning method EIB for better transfer learning results under the IB principle, which is flexible to suit different kinds of data and can be merged into domain adaptation algorithms as a regularizer.
- In the study of SG, we provide a tighter SG upper bound, which serves as a theoretical guarantee for DIB and further develop the generalization theory in the IB framework.
- Comprehensive simulations and experiments validate our theoretic results and demonstrate that EIB outperforms IB and DIB in many cases.

## 2. Problem Formulation

In domain adaptation, a scenario of transductive transfer learning, we have labeled training data from a source domain, and we wish to learn an algorithm on the training data that performs well on a target domain [27]. The learning task is the same on the two domains, but the population distribution of the source domain and the target domain are different. Specifically, the source and the target instances come from the same instance space $\mathcal{X}$, and the same instance corresponds to the same label, even if they lie in different domains. However, the population instance distribution on the source and the target domain are different, denoted as $p(X)$ and $q(X)$. (We use "instance" (X) to distinguish from "label" (Y) and "example" (X,Y) and use "population distribution" ($p$ or $q$) to distinguish from "empirical distribution" ($\hat{p}$ or $\hat{q}$)). Assume that there exists a generic feature space so it can be utilized to transfer knowledge between different domains. That is to say, both the IB and DIB methods involve an encoder with a transition probability $p(t|x)$ to convert instance $\mathcal{X}$ to an intermediate feature $T \in \mathcal{T}$. The induced marginal distribution is then denoted as $p(t)$ and $q(t)$ for the source and target domains, respectively. The IB and DIB methods also have a decoder or classifier $h \in \mathcal{H}$ to map from the feature space $\mathcal{T}$ to the label space $\mathcal{Y}$. $p(y|x)$ denotes the ground-truth labeling function and $p(y|t) = \sum_x p(y|x)\frac{p(t|x)p(x)}{\sum_{x'} p(t|x')p(x')}$ is a labeling function induced by $p(y|x)$ and $p(t|x)$. In classification scenarios, deterministic labels are commonly used. Therefore, we define the deterministic ground-truth-induced labeling function as $f : \mathcal{T} \rightarrow \mathcal{Y}, f(t) = argmax_y p(y|t)$. If the maximum probability is not unique, i.e., $\{y_1, \ldots, y_n\} = argmax_y p(y|t)$, randomly choose a $y_{i,i\in\{1...n\}}$ as the output. With the above definitions, the expected error on the source domain could be written as $\epsilon_S(h) = E_{t\sim p(t)}I_{\{f(t)\neq h(t)\}}$, where $I_{\{\cdot\}}$ is the indicator function. Similarly, the expected error on target domain is $\epsilon_T(h) = E_{t\sim q(t)}I_{\{f(t)\neq h(t)\}}$. Then, our problem could be formulated as follows: When the IB and DIB methods are trained on the source domain, which method achieves a lower target domain expected error $\epsilon_T(h)$?

According to previous work [26], the target domain error could be decomposed to three parts, as shown in the following theorem. A detailed proof is provided in Appendix A.1.

 

**Theorem 1** (Target Error Decomposition). *Suppose that $h^*$ is the classifier in $\mathcal{H}$, which minimizes the sum of the expected error in two domains, i.e., $h^* = argmin_{h \in \mathcal{H}} \epsilon_S(h) + \epsilon_T(h)$. Let us denote $\epsilon_S(h^*) + \epsilon_T(h^*) \triangleq \lambda_S + \lambda_T = \lambda$. Then, for a classifier $h$, we have*

$$\epsilon_T(h) \leq \hat{\epsilon}_S(h) + \delta_S(h) + d_{\mathcal{H} \Delta \mathcal{H}}(p(t), q(t)) + \lambda, \tag{1}$$

*where $\hat{\epsilon}$ is the empirical error; $\delta_S(h)$ is the source generalization error gap, i.e., $\delta_S(h) = |\hat{\epsilon}_S(h) - \epsilon_S(h)|$; and $d_{\mathcal{H} \Delta \mathcal{H}}(p(t), q(t))$ is the RD defined by the $\mathcal{A}$-distance, i.e., $d_{\mathcal{H} \Delta \mathcal{H}}(p(t), q(t)) = \sup_{h \in \mathcal{H}} |E_{t \sim p(t)} I_{\{h^*(t) \neq h(t)\}} - E_{t \sim q(t)} I_{\{h^*(t) \neq h(t)\}}|$.*

Please note that we assume that the ground-truth labeling function on the two domains remains the same, which is common, as used in many previous studies. If there is a conditional shift, i.e., a distance between $f_S$ and $f_T$, where $f_S$ and $f_T$ are the deterministic ground-truth-induced labeling function on the source and target domains, respectively, as defined in Zhao et al. [28], the above decomposition is not valid any more. The assumption is reasonable since the conditional shift phenomenon is not observed in our experiments.

According to the above theorem, we need to minimize the following three terms to achieve a low expected error, i.e., the training error on source domain, and SG and RD between the marginal distributions on the two domains (RD). Assume that both the IB and DIB methods have the ability to achieve comparable small source training errors, so we focus on SG and RD to compare the two methods.

## 3. Main Results

### 3.1. Source Generalization Study of IB and DIB

In this subsection, we study the generalization of IB and DIB on the source domain. The generalization error is used to quantify the degree to which a supervised machine learning algorithm may overfit the training data. The generalization error gap in statistical learning theory is defined as the expected difference between the population risk and the empirical risk. Russo and Zou [29], Xu and Raginsky [30] provided a generalization upper bound with mutual information of the input and output. Sefidgaran et al. [31,32] further related the mutual information with two other studies of generalization error, i.e., compressibility and fractal dimensions, providing a unifying framework for the three directions of studies.

In the theoretic study of generalization and adversarial robustness of IB, however, the measure is not exactly the same for the convenience of illustrating the characteristic of IB [8,9]. Specifically, in Section 4.1 in Shamir et al. [8], the classification error is proven to be exponentially upper bounded by $-I(Y; T)$, indicating that $I(Y; T)$ is a measure of performance. Therefore, the difference between the population performance $I(Y; T)$ and empirical performance $\hat{I}(Y; T)$ is used to measure the ability of generalization. Shamir et al. [8] provided the following generalization upper bound, which is relevant to the mutual information $I(X; T)$.

**Theorem 2** (The previous generalization bound). *Denote the empirical estimates by $\hat{\cdot}$. For any probability distribution $p(x, y)$, with a probability of at least $1 - \delta$ over the draw of the sample of size $m$ from p(x,y), we have that for all $T$,*

$$|I(Y; T) - \hat{I}(Y; T)| \leq$$
$$\sqrt{\frac{B_0 \log(|\mathcal{Y}|/\delta)}{m}} \left( B_1 \log(m) \sqrt{|\mathcal{T}|I(X; T)} + B_2 |\mathcal{T}|^{3/4} (I(X; T))^{1/4} + B_3 \hat{I}(X; T) \right) \tag{2}$$

*where $B_0$ is a constant and $B_1$, $B_2$, $B_3$ only depend on $m$, $\delta$, $|\mathcal{Y}|$, $\min_x p(x)$, and $\min_y p(y)$, where* min *refers to the minimum value other than 0.*

However, this upper bound cannot be directly used to compare IB and DIB because both IB and DIB minimizes $I(X; T)$. Moreover, the regularizer of DIB: $H(T) = I(X; T) + H(T|X)$ further minimizes the term $H(T|X)$. However, it is not clear whether it may bring any advantage from the previous theoretical result. To tackle this problem, we prove a new generalization bound in this paper, as shown in the following theorem.

**Lemma 1.** *If $X \in [0,1]$, $EX = \mu$, then $Var(X) \leq (1 - \mu)\mu$.*

**Lemma 2.** *If $\sum x = 1$, $\forall 0 \leq x \leq 1$, then $\sum_x -\sqrt{x(1-x)}\, ln(\sqrt{x(1-x)}) \leq \sum_x -\sqrt{x}\, lnx$.*

**Theorem 3** (Our generalization bound). *For any probability distribution $p(x, y)$, with a probability of at least $1 - \delta$ over the draw of the sample of size $m$ from $p(x, y)$, we have that for all $T$,*

$$
|I(Y;T) - \hat{I}(Y;T)| \leq
$$
$$
\frac{1}{\sqrt{m}}\left((C_1 + C_3)\sqrt{|\mathcal{T}| - 1} + C_2 H(T) + C_4 H(T|Y) + C_5\sqrt{(\log|\mathcal{T}| - \hat{H}(T|Y))\hat{H}(T|Y)}\right) \tag{3}
$$

*where $C_1, C_2, C_3, C_4, C_5$ only depend on $\delta$, $|\mathcal{Y}|$, $\min_{x,y} p(x|y)$, $\min_{t,y} p(t|y)$, $\min_x p(x)$, and $\min_t p(t)$.*

**Proof.** Here, we show a proof sketch to help us understand how our bound is different from the previous one. The complete proof is in the appendix. Similarly to the previous proof, SG is firstly divided into three parts.

$$
|I(Y;T) - \hat{I}(Y;T)| \leq
$$
$$
|H(T) - \hat{H}(T)| + \Big|\sum_y p(y)\big(H(T|y) - \hat{H}(T|y)\big)\Big| + \Big|\sum_y (p(y) - \hat{p}(y))\hat{H}(T|y)\Big| \tag{4}
$$

Denote $\Delta_1 = |H(T) - \hat{H}(T)|$, $\Delta_2 = \big|\sum_y p(y)(H(T|y) - \hat{H}(T|y))\big|$, $\Delta_3 = \big|\sum_y (p(y) - \hat{p}(y))\hat{H}(T|y)\big|$. Then, we can summarize the previous proof and our proof in Figure 1.

| The previous proof | | Our proof | |
|---|---|---|---|
| $\Delta_1 + \Delta_2 \leq 2\sum_t \phi(B\sqrt{|\mathcal{X}|V_x(p(t|x))})$ (5) | | $\Delta_1 \leq \sum_t \phi(\frac{B}{\sqrt{min_x p(x)}}\sqrt{Var_X(p(t|x))})$(12) $\leq \sum_t \phi(\frac{B}{\sqrt{min_x p(x)}}\sqrt{(1-p(t))p(t)})$(13) | |
| $\leq 2\sum_t \phi(\frac{2B}{min_x p(x)}\|\frac{p(x|t)}{p(x)} - 1\|_1)$ (6) | | $\leq \frac{C_1}{\sqrt{m}}\sqrt{|T| - 1} + \frac{C_2}{\sqrt{m}}H(T)$ (14) | |
| $\leq 2\sum_t \phi(\frac{2B}{min_x p(x)}\sqrt{2log(2)}\sqrt{D_{KL}[p(x|t)\|p(x)]})$ (7) | | $\Delta_2 \leq \sum_y p(y)\sum_t \phi(\frac{B}{\sqrt{min_x p(x|y)}}\sqrt{Var_{X|Y}(p(t|x))})$ (15) | |
| $\leq B(B_1 log(m)\sqrt{|\mathcal{T}|I(X;T)} + B_2|\mathcal{T}|^{3/4}I(X;T)^{1/4})$ (8) | | $\leq \sum_y p(y)\sum_t \phi(\frac{B}{\sqrt{min_{x,y} p(x|y)}}\sqrt{(1-p(t|y))p(t|y)})$ (16) $\leq \frac{C_3}{\sqrt{m}}\sqrt{|T| - 1} + \frac{C_4}{\sqrt{m}}H(T|Y)$(17) | |
| $\Delta_3 \leq B\sqrt{|\hat{\mathcal{Y}}|V_y(\hat{H}(T|y))}$ (9) | | $\Delta_3 \leq \frac{B}{\sqrt{min_y p(y)}}\sqrt{Var_Y(\hat{H}(T|y))}$ (18) | |
| $\leq B(1 + 1/\sqrt{y})\|\hat{H}(T) - \hat{H}(T|y)\|_1$ (10) | | $\leq \frac{B}{\sqrt{min_y p(y)}}\sqrt{(log|T| - \hat{H}(T|y))\hat{H}(T|y)}$ (19) | |
| $\leq \frac{2B}{min_y p(y)}\hat{I}(X;T)$ (11) | | $= \frac{C_5}{\sqrt{m}}\sqrt{(log|T| - \hat{H}(T|Y))\hat{H}(T|Y)}$ (20) | |

**Figure 1.** The previous proof vs. our proof.

The main idea of the previous proof is to bound these parts by $\phi$ functions of sample variances in (5) and (9), where $V_x(p(t|x)) \triangleq \frac{1}{|\mathcal{X}|}\sum_x (p(t|x) - \frac{1}{|\mathcal{X}|}\sum_x p(t|x))^2$, and $\phi(x)$ is defined by (A1). The variances are then related to the $L_1$ distances in (6) and (10); KL-divergence in (7); and lastly, mutual information in (8) and (11).

We also use the idea of bounding by $\phi$ functions of variances, as shown in (12), (15), and (18). However, the variance is an exact one, instead of the previous estimation on the samples, i.e., $Var_X(p(t|x)) \triangleq E_{p(x)}[p(t|x) - E_{p(x)}[p(t|x)]]^2$. Consequently, we can obtain a tighter bound. Specifically, we first use Lemma A4 to convert the variance to the expectation in (13), (16), and (19). Then, with the help of Lemma A5, we turn the $\phi$ functions into entropy, like from (13) to (14). With (14), (17), and (20), we finish the proof of Theorem 3. □

We can see that this bound is similar to the previous one, but the new bound is related to the regularization term $H(T)$, $H(T|Y)$ and $\hat{H}(T|Y)$, instead of the mutual information $I(X;T)$. Note that $H(T|Y)$ and $\hat{H}(T|Y)$ are closely related by $|H(T|Y) - \hat{H}(T|Y)| \leq$ (17) + (20). Additionally, $H(T)$ is equivalent to $H(T|Y)$ as a regularizer because $L_{DIB} = H(T) - \beta I(Y;T) = H(T|Y) - (\beta - 1)I(Y;T)$. With these two observations above, this bound is directly dependent on the entropy $H(T)$, i.e., the regularizer of DIB. That is to say, compressing the entropy of the representation is beneficial to reducing SG.

Now we will illustrate why DIB generalizes better. Intuitively, since DIB directly minimizes $H(T)$ while IB only minimizes a part of them, i.e., $I(X;T)$, DIB may have a better generalization than IB. Owing to the lack of close form solutions, we cannot explicitly see the difference between IB and DIB with respect to the concerning information quantities. However, we can solve the IB and DIB problems through a self-consistent iterative algorithm, with $\alpha = 0$ or $1$ in Appendix B.1. The IB and DIB solutions are shown on the information plain in Appendix C with $\alpha = 0$ or $1$, showing that the IB solutions have larger $H(T)$ than DIB solutions, and thus, DIB generalizes better in the sense of Theorem 3 in our paper.

Furthermore, our bound is tighter than the previous one. We provide a comparison of the order in Appendix A.2, which shows that the IB bound is $\sqrt{|X|}$ times larger than the DIB bound for the first two terms and $\sqrt{|Y|}$ times larger for the third term. There is also an empirical comparison in Section 4.1.2. Moreover, Theorems 2 and 3 require some constraints for sample size. They are summarized in the Appendix A.2 too. The empirical results show that our bound requires a smaller sample size.

To sum up, we provide a tighter generalization bound in the IB framework, which serves as a theoretical support for DIB's generalization performance. Experiments on MNIST will validate this theoretical result in Section 4.2.1.

### 3.2. Representation Discrepancy of IB and DIB

In this section, we compare the RD of IB and DIB. According to the target error decomposition theorem in Section 1, RD is measured by $\mathcal{H}\Delta\mathcal{H}$-distance, which is however difficult for direct computation because of the complex hypothesis space $\mathcal{H}$. To remove the dependence of $\mathcal{H}$, we propose to bound RD by a pair-wise $L_1$ distance.

To start with the simplest case, assume that the sample sizes on the source and target domains are both $m$, and the samples on two domains have a one-to-one correspondence (i.e., semantically close to each other). Then, RD can be bounded by the following pair-wise $L_1$ distance.

**Proposition 1.** *The distance of overall representations on the source and target domains is bounded by the distance of individual instance representations.*

$$d_{\mathcal{H}\Delta\mathcal{H}}(p(t), q(t)) \leq \frac{1}{m} \sum_{(x_S, x_T) \in \mathcal{CP}} \|p(t|x_S) - p(t|x_T)\|_1 + \epsilon, \tag{21}$$

*where $\mathcal{CP}$ stands for the set of all correspondence pairs and $\epsilon = 2\sqrt{|\mathcal{T}|}\frac{2+\sqrt{2\log(1/\delta)}}{\sqrt{m}}$, which is small when the sample size is large.*

In fact, Proposition 1 is valid for any $(x_S, x_T)$ pair, so there can be different upper bounds. To obtain the lowest upper bound, we define correspondent pairs such that $\sum_{(x_S, x_T) \in \mathcal{CP}} \|p(t|x_S) - p(t|x_T)\|_1$ is the lowest, i.e., $\mathcal{CP} = \{(x_{S,i}, x_{T,i})_{i=1...m} = \arg\min \sum_{i=1}^{m} \|p(t|x_{S,i}) - p(t|x_{T,i})\|_1 | \{x_{S,i}\}_{i=1...m} = \mathcal{X}_S, \{x_{T,i}\}_{i=1...m} = \mathcal{X}_T\}$.

We parameterize $p(t|x)$ to be a $d$-dimensional Gaussian with a diagonal covariance matrix, which is usually assumed in some variational methods such as Alemi et al. [15]. Compared with IB, DIB additionally reduces $H(T|X)$, so its $p(t|x)$ are almost deterministic. Therefore, the variance in $p_{DIB}(t|x)$ in each dimension are significantly smaller than that of $p_{IB}(t|x)$. On the other hand, since the only difference between IB and DIB is altering $H(T|X)$ and the entropy of Gaussian random variable is only dependent on its variance, the expectations of $p_{IB}(t|x)$ and $p_{DIB}(t|x)$ are comparable. Therefore, we need to find how the variances affect RD.

Consider the representations of the instances from the source and target domains in the same model. $\forall (x_S, x_T) \in \mathcal{CP}$, since $x_S$ and $x_T$ are semantically close, the expectations of their representations are close. Additionally, the discrepancy between their variances are small compared to the discrepancy in the representation variances between IB and DIB, so we can neglect them. Therefore, denote that $p(t|x_S) \sim N(\boldsymbol{\mu}_1, \Sigma)$, $p(t|x_T) \sim N(\boldsymbol{\mu}_2, \Sigma)$, $\Sigma = diag(\sigma_1^2, \ldots, \sigma_d^2)$; then,

$$\|p(t|x_S) - p(t|x_T)\|_1 = \Pi_{i=1}^d (4\Phi(\frac{|\mu_{1i} - \mu_{2i}|}{2\sigma_i}) - 2) \tag{22}$$

where $\Phi$ is the cumulative distribution function of a standard Gaussian distribution.

As discussed above, compared with IB, DIB has significantly smaller variances in the representations and $\mu_1 - \mu_2$ are comparable for IB and DIB. Therefore, the term $\frac{|\mu_{1i} - \mu_{2i}|}{2\sigma_i}$ for IB is remarkably smaller than DIB. With $\Phi$ monotonically increasing, $\|p(t|x_S) - p(t|x_T)\|$ for IB is smaller than that for DIB. Figure 2 provides an intuitive understanding about how randomness helps reduce RD. Suppose that the blue and red lines are $p(t|x_S)$ and $p(t|x_T)$, where $(x_S, x_T) \in \mathcal{CP}$. We can see clearly that their $L_1$ distance drops with the growth in variances. Moreover, because the derivative of $\phi$ monotonically decreases on $[0, +\infty]$, when $\mu_2 - \mu_1$ is smaller, the difference between IB and DIB will be larger. This phenomenon is also found in simulations; see Sections 4.1.1 and 4.1.3.

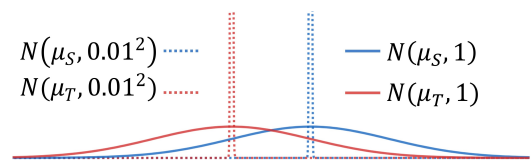

**Figure 2.** The effect of randomness on RD.

When the sample size on the two domains are different and the "pair-wise" correspondence does not hold, we can take the correlated instances in the two domains as a general form of correspondent pairs; then, the above comparison result is also valid. Details are found in Appendix A.3.

Please note that the assumption of correspondent pairs is reasonable. In transfer learning, it is widely assumed that there exists shared (high level) features and common distributions of representation on the two domains. The correspondent pairs can be viewed as the instance pairs with the closest distributions of representations from the two domains.When the distributions are distinct, feature alignment is usually implemented in practice.

According to the above theoretical results, we can obtain the following comparisons between IB and DIB, with respect to the three terms for the target error decomposition, as illustrated in Table 1. Clearly, there is a trade-off between IB and DIB, in terms of SG and RD.

**Table 1.** A comparison of IB and DIB.

| Decomposition Terms | IB vs. DIB |
| --- | --- |
| Source Training Error | IB $\approx$ DIB |
| Source Generalization Gap (SG) | IB > DIB |
| Representation Discrepancy (RD) | IB < DIB |

### 3.3. Elastic Information Bottleneck Method

From the previous results, IB and DIB still have room for improvement, respectively, on SG and RD. To obtain a Pareto optimal solution under the IB principle, we propose a new bottleneck method as follows, namely EIB. Please note that the generalized IB objective function in [7] is the same as the objective function of elastic IB, but they are derived from different perspectives. The generalized IB is constructed for the convenience of solving the DIB problem, while EIB is proposed to balance SG and RD.

**Definition 1** (Elastic Information bottleneck). *The objective function of the elastic information bottleneck method is as follows:*

$$\min L_{EIB} = (1 - \alpha)H(T) + \alpha I(X; T) - \beta I(Y; T), 0 \le \alpha \le 1, \beta > 1 \tag{23}$$

We can see that EIB is a linear combination of IB and DIB, which covers IB and DIB as special cases. Specifically, EIB reduces to IB when $\alpha = 1$ and DIB when $\alpha = 0$. Since IB and DIB perform better than the other one on SG or RD, respectively, a linear combination may lead to better target performance . In fact, the global optimal solution of the linear combination of the objectives is in the Pareto optimal solution set. Therefore, by adjusting $\alpha$ in $[0, 1]$, we can obtain a balance between SG and RD and achieve a Pareto frontier within the IB framework.

As a bottleneck method similar to IB, its optimal solution can be calculated by iterative algorithms when the joint distribution of instances and labels are known. The algorithm and EIB's optimal solutions on information planes are provided in Appendix B. However, the iterative methods are intractable when the data are numerous and high-dimensional. Therefore, we use the variational approaches similar to the method in Fischer [2] to optimize IB, DIB, and EIB objective functions by neural networks. Assume the representation is a k-dimension Gaussian distribution and all dimensions are independent. The network contains an encoder $p(t|x)$, a decoder $q(y|t)$, and a backward encoder $b(t|y)$. To be specific, the encoder is an MLP which outputs the K-dimensional expectations $\boldsymbol{\mu_1}$ and K-dimensional diagonal elements of covariance matrix $\boldsymbol{\sigma_1}$ and then yield the Gaussian representation with the reparameterization trick [15]. The decoder is a simple logistic regression model. The backward encoder is an MLP that uses the onehot encoding of the classify outcome as an input and the K-dimensional expectations $\boldsymbol{\mu_2}$ and K-dimensional diagonal elements of covariance matrix $\boldsymbol{\sigma_2}$ as an output. The loss function of variational EIB is as follows:

$$Loss_{EIB} = \int p(t|x)p(x,y)\big(\log \frac{p^{\alpha}(t|x)}{b(t|y)} - \tilde{\beta}\log q(y|t)\big)\mathrm{d}x\mathrm{d}t\mathrm{d}y, 0 \le \alpha \le 1, \beta > 0.$$

$$Loss_{EIB} \approx \frac{1}{m}\sum_{n=1}^{m}\big[\sum_{j=1}^{k}\big[\frac{(1-\alpha)}{2}ln(2\pi) - \alpha ln\sigma_{1,j,n} + ln\sigma_{2,j,n} - \frac{\alpha}{2}$$

$$+ \frac{\mu_{1,j,n}^2 + \mu_{2,j,n}^2 - 2\mu_{1,j,n}\mu_{2,j,n} + \sigma_{1,j,n}^2}{2\sigma_{2,j,n}^2}\big] + \tilde{\beta}CE[softmax(y_n)||q(y_n|t_n)]\big], \tag{24}$$

where $CE$ is the cross entropy loss.

Our simulations and real data experiments show that EIB outperforms IB and DIB. It is also worth noticing that EIB can be plugged into previous transfer learning algorithms to expand its application. An example is given in the next section.

## 4. Experiments

In this section, we evaluate IB, DIB, and EIB by both toy data simulations and real data experiments.

### 4.1. Toy Data Simulations

We conduct simulations to study the performance of EIB in the transfer learning scenario, a comparison of RD between IB and DIB, and the comparison between our generalization bound and the previous one.

### 4.1.1. Performance of EIB in Transfer Learning

We design a toy binary classification problem, where instance $X \in \{0,1\}^{10}$ and label $Y \in \{0,1\}$. Define $X_0 = [1,1,1,1,1,0,0,0,0,0]$, $X_1 = [0,0,0,0,0,1,1,1,1,1]$, $Y_0 = 0$, and $Y_1 = 1$. First, in the source and target datasets, half of the examples is $(X_0, Y_0)$ and the other half is $(X_1, Y_1)$. Then we introduce random noise into the instances in the following way. We define the reverse digit operation as first choosing one digit in the instance and then reversing the digit (changing from 0 to 1 or the opposite direction). For each instance, we perform the reverse digit operation $\lfloor N \rfloor$ times (round down function), where $N$ is a real-valued uniform random variable $N \sim U[0, R]$. $R$ is named as the noise level because as R increases, more instances are very different from $X_0$ and $X_1$ in the dataset. As a result, we can adjust the similarity between two domains by modifying the parameter $R$. Lastly, we set the $R \in (1,3)$ for the source domain and $R = 3$ for the target domain so that we obtain the toy data in a transfer learning scenario.

We test our EIB model on the toy data. The parameter $\alpha$ ranges in [0,1], indicating different EIB models. The other parameter $\beta$ is chosen to be $10^4$ for $R = 2, 1.5, 1.433$ and $\beta = 5 \times 10^3$ for $R = 1.375, 1.25$ in order to discriminate the model performance and to obtain high accuracy. The results are shown in Table 2. From the results, we can see that EIB outperforms IB and DIB when $R = 1.5, 1.433, 1.375, 1.25$. To be more specific, when the two domains become more similar, i.e., the noise level $R$ of the source domain becomes closer to $R = 3$ of the target domain, $\alpha$ of the best model changes from 0 to 1, i.e., IB gradually becomes more advantageous. This is in accordance with our theory that when the two domains are similar, the effect of RD is apparent and IB has more advantages in transfer learning. This result also provides an empirical rule to tune the parameter $\alpha$.

**Table 2.** Accuracy of elastic information bottleneck (EIB) on simulated data with different noise levels R. The highest accuracy under each noise levels are marked in bold.

| $\alpha$ | R = 2 | R = 1.5 | R = 1.433 | R = 1.375 | R = 1.25 |
|----------|-------|---------|-----------|-----------|----------|
| 0 | 98.85 | 97.20 | 98.55 | 97.65 | 98.30 |
| 0.1 | 98.30 | 97.40 | 98.70 | 97.30 | **98.35** |
| 0.2 | 98.45 | 97.40 | **98.75** | **98.15** | 98.05 |
| 0.3 | 98.45 | 97.70 | 98.65 | 97.85 | 98.20 |
| 0.4 | 98.50 | 97.75 | 98.60 | 97.45 | 98.20 |
| 0.5 | 99.10 | 97.40 | 98.65 | 97.55 | 98.25 |
| 0.6 | 99.15 | 97.45 | 98.55 | 97.40 | 98.25 |
| 0.7 | 98.60 | 97.65 | 98.60 | 97.35 | 98.25 |
| 0.8 | 98.65 | 97.55 | 98.60 | 97.25 | 98.00 |
| 0.9 | 98.60 | **98.53** | 98.65 | 97.20 | 98.10 |
| 1.0 | **99.25** | 97.75 | 98.65 | 97.50 | 97.90 |

### 4.1.2. Our Generalization Bound vs. the Previous One

In this simulation, we compare the DIB generalization bound (ours) and the IB generalization bound [8] in terms of number size and required sample size.

Assume that the data, representations, and labels are discrete, i.e., $|\mathcal{X}| = 3, |\mathcal{T}| = 2$, $|\mathcal{Y}| = 2$, $\delta = 0.1$. The sample size increases exponentially from $10^1$ to $10^6$. The bound and error rate are computed as follows. First, we generate distributions $p(x,y)$, $p(t|x)$ and sample an empirical distribution $\hat{p}(x,y)$. To compare the bounds in a general case, $p(x,y)$ is randomly valued by numbers generated from a uniform or a normal distribution and the value of $p(t|x)$ is also randomly generated from a uniform distribution or a normal distribution. $\hat{p}(x,y)$ is the empirical probability of $p(x,y)$. Second, we determine whether the constraints for the bound in Appendix A.3 are met. If so, we calculate the value of the bound. If not, we record the number of times the constraints are not satisfied. Third, we repeat this process for 100 times to obtain the rate of constraint violations (error rate) and two average bound values. Finally, we repeat the previous three steps under different sample sizes.

Figure 3 (left) shows that our bound consistently needs less samples than the previous one to satisfy the constraints. Moreover, the more uniform the distribution is, the less samples are needed to satisfy the constraints, which is obvious from the formula of constraints. Figure 3 (right) shows that the generalization bound decreases as the sample size increases, and the DIB generalization bound is always smaller than the IB generalization bound.

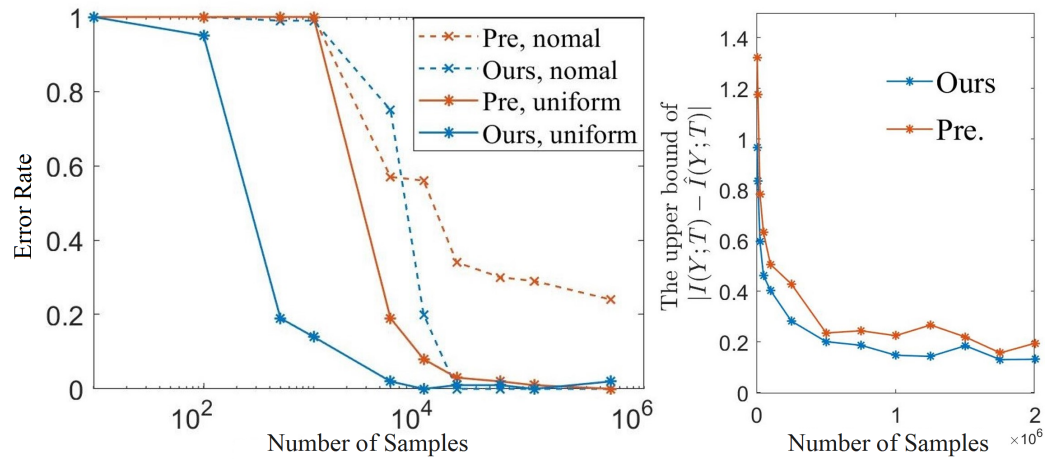

**Figure 3.** Pre. denotes the previous bound. **Left:** Comparisons of the two bounds with respect to the constraint error as the sample size grows. **Right:** Comparisons of the two bounds with respect to the tightness when samples are sufficient.

### 4.1.3. Comparison of RD between IB and DIB

To support the claims about RD, we approximate RD by using a classifier to predict which domain the representation sample $t$ comes from, which is proven to be an adequate approximation of RD in [26].

The data generation, model, and parameters (R,$\beta$) are consistent with the simulation in Section 4.1.1. The data are generated two times under different noise levels R and are trained by EIB models with $\alpha = 0$ or 1. Then, five samples are drawn from the Gaussian representations $p(t|x)$ for each instance x. If x comes from the source domain, we label the samples as positive; otherwise, we label it as negative. Then, we train a linear classifier to classify the samples. Each classifier is trained 20 times. The average error rate is shown in Table 3. The standard error of the mean (SEM) is in parentheses.

**Table 3.** The average error rate of classification between representation samples on two domains with difference noise levels. A larger error rate indicates smaller RD. The highest error rate under each noise levels are marked in bold.

| R | 2 | 1.5 | 1.433 | 1.375 | 1.25 |
|---|---|---|---|---|---|
| $\beta$ | $10^4$ | $10^4$ | $10^4$ | $5 \times 10^3$ | $5 \times 10^3$ |
| DIB ($\alpha = 0$) | 43.65% $\pm$ 0.20% | 28.05% $\pm$ 0.15% | 28.14% $\pm$ 0.35% | 28.99% $\pm$ 0.71% | **27.72% $\pm$ 0.50%** |
| IB ($\alpha = 1$) | **45.46% $\pm$ 0.26%** | **28.58% $\pm$ 0.18%** | **28.54% $\pm$ 0.35%** | **32.84% $\pm$ 1.25%** | 27.46% $\pm$ 0.35% |

The results show that IB has a larger error rate than DIB in most cases, indicating that IB has a smaller RD, which is consistent with the main result in Section 3.2. Furthermore, the gap between IB's and DIB's error rates becomes larger with the growth of R, which validates that when two domains become more similar, IB's advantage over RD becomes more significant, as is claimed in Sections 3.2 and 4.1.1. This trend does not seem to be consistent between $R = 1.433$ and $R = 1.375$ because $\beta$, which is chosen to minimizes the error rate, is different.

### 4.2. Real Data Experiments

Similar to IB, EIB can also be utilized in many representation learning algorithms as a regularizer. As an example, we combine EIB with DFA-MCD in Wang et al. [33], which is an adversarial domain adaptation algorithm with feature alignment. We replace the features in DFA-MCD with Gaussian representations and add a backward encoder after the source classifier as in variational EIB. Since the feaures are already Gaussian, we remove the KLD regularizer in DFA-MCD, which is the KL-divergence penalty between source domain feature and a Gaussian prior. Then, we substitute cross entropy loss with EIB loss. The adversary training and feature alignment designs are retained.

We test the model on a common transfer learning task, where the source dataset is MNIST [34] and the target dataset is USPS [35]. We only use 2/55 of the training data, and $\lambda = 15$, $\beta = 10^9$, other parameters remain the same as in Wang et al. [33]. The experiment is repeated with six random seeds, and the results are shown in Figure 4 (left). For each box, the central mark indicates the median, and the bottom and top edges of the box indicate the 25th and 75th percentiles, respectively. The whiskers extend to the most extreme data points not considered outliers, and the outliers are plotted individually using the '+' symbol. It is automatically generated by the boxplot function in matlab. We can see that the model with EIB performs better than IB and DIB in most cases, e.g., $\alpha = 0.2, 0.3, 0.4, 0.6, 0.7, 0.8$, and some of them beats the DFA-MCD baseline model, which suggests that EIB works as an effective regularizer for transfer learning. Please note that this example is utilized to demonsrate EIB can be combined with domain adaptation algorithms and perform better than IB and DIB, so we simply inherit the structure and hyper-parameters of the original networks in Wang et al. [33]. Further parameter tuning can be conducted to achieve better experimental results.

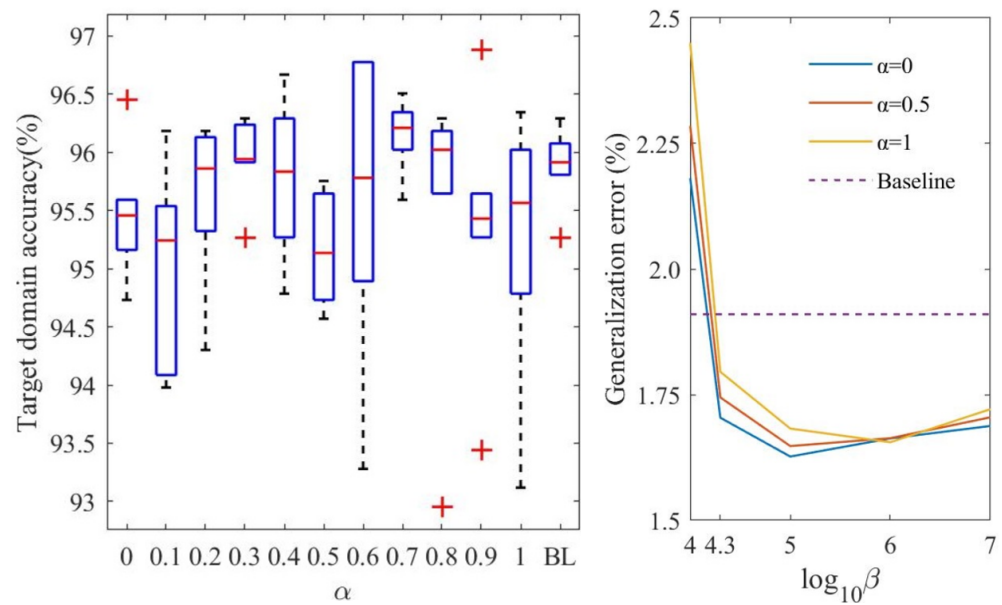

**Figure 4. Left:** Target domain accuracy by EIB in transferring MNIST to USPS. BL (baseline) is the DFA-MCD method. **Right:** SG on MNIST by EIB.

#### 4.2.1. Source Generalization Analysis

We use variational EIB on MNIST to compare SG of IB and DIB. The networks were trained for 100 epochs and converge at about 60 epochs. The average generalization gap is calculated by the mean discrepancy of the training error and the testing error of 25 epochs with the least testing error. We randomly initialize the network 14 times and utilize EIB models with $\alpha = 0$ (DIB), $\alpha = 0.5$, $\alpha = 1$ (IB) under the same initialization. The results are shown in Figure 4 (right) and Table 4. Baseline is a model without a regularizer i.e., $\beta = +\infty$ in EIB. First, let us analyze the results in terms of $\beta$. When $\beta$ is small, models over-compress the representations so that the error rate is large. When $\beta$ is large, the weight of regularization term in objective function is small so the three models' performance become close. When $\beta = 10^5$, the three models have the best accuracy. Then, in terms of $\alpha$, with the decrease in $\alpha$, the generalization gap becomes smaller. When $\beta = 10^4, 5 \times 10^4, 10^5$, the $p$ value of the $t$-test on "DIB's SG < IB's SG" is less than 0.05, indicating that the source generalization error of DIB is statistically significantly smaller than that of IB. (The $p$ value is a term in statistical test that is defined as the probability of obtaining a test statistic as extreme or more extreme than the test statistic of actual observations if the null hypothesis is true. Usually, if $p$ value < 0.05, the null hypothesis should be rejected and the result is considered statistically significant) This validates that DIB generalizes better than IB.

**Table 4.** T-test on "DIB's SG < IB's SG". Statistically significant results are marked in bold.

| $\beta$ | $10^4$ | $5 \times 10^4$ | $10^5$ | $10^6$ | $10^7$ | $10^8$ |
|---|---|---|---|---|---|---|
| **$p$ value** | **0.021** | **0.020** | **0.042** | 0.66 | 0.085 | 0.39 |

## 5. Conclusions and Future Work

This work studies the two objective functions of the information bottleneck principle. The motivation comes from our theoretical analysis that neither IB nor DIB is the optimal solution in terms of the generalization ability under the transfer learning scenario. Specifically, we theoretically analyze SG and RD of IB and DIB, and find that there is a trade-off between them. To tackle this problem, we propose a new method, EIB, to interpolate IB and DIB. Consequently, EIB can not only achieve better transfer learning performances but also be plugged into existing domain adaptation methods as a regularizer to suit different kinds of data, which have been shown by our simulations and real data experiments.

We believe that our results take an important step towards understanding different information bottleneck methods and provide some insights into the design of stronger deep domain adaptation algorithms. We qualitatively suggest choosing $\alpha$, but in practice, when the distance of the two domain is fixed, the optimum still needs careful tuning. Therefore, how to choose the best parameters $\alpha$ and $\beta$ in EIB remain questions that we plan to study in the future.

**Author Contributions:** Methodology, Y.N. and Y.L.; Formal analysis, Y.N.; Software, A.L.; Writing—original draft preparation, Y.N.; Writing—review and editing, Y.L.; Supervision, Y.L. and Z.M. All authors have read and agreed to the published version of the manuscript.

**Funding:** This work is supported by National Key R&D Program of China No. 2021YFF1201600, Vanke Special Fund for Public Health and Health Discipline Development, Tsinghua University (No. 2022-1080053), and Beijing Academy of Artificial Intelligence (BAAI).

**Institutional Review Board Statement:** Not applicable.

**Informed Consent Statement:** Not applicable.

**Data Availability Statement:** The source code is available at https://github.com/nyyxxx/elastic-information-bottleneck, accessed on 20 July 2022.

**Conflicts of Interest:** The authors declare no conflicts of interest.

## Appendix A. Missing Proofs

*Appendix A.1. Target Error Decomposition*

**Lemma A1.** *The triangle inequality for classification error holds.* $\epsilon(f_1, f_2) \leq \epsilon(f_1, f_3) + \epsilon(f_2, f_3)$, *where* $\epsilon(f_1, f_2) = E_z I_{\{f_1(z) \neq f_2(z)\}}$.

**Proof.** $A \triangleq \{f_1 \neq f_2\}, B \triangleq \{f_1 \neq f_3\}, C \triangleq \{f_2 \neq f_3\}$. Obviously, $B^c \cap C^c \subseteq A^c$. This leads to the inverse proposition $B \cup C \supseteq A$. Then, $I_B + I_C \geq I_{B \cup C} \geq I_A$. Add expectation $E_z I_B + E_z I_C \geq E_z I_A$. With the definition of the classification error, the triangle inequality is proven. □

**Proof of Theorem 1.**

$$\epsilon_T(h) \leq \epsilon_T(h^*, f_T) + \epsilon_T(h, h^*) = \lambda_T + E_{z \sim q(t)} I_{\{h^*(z) \neq h(z)\}}$$

$$= \lambda_T + E_{z \sim p(t)} I_{\{h^*(z) \neq h(z)\}} + (E_{z \sim q(t)} I_{\{h^*(z) \neq h(z)\}} - E_{z \sim p(t)} I_{\{h^*(z) \neq h(z)\}})$$

$$\leq \lambda_T + \epsilon_S(h, h^*) + d_{\mathcal{H}\Delta\mathcal{H}}(p(t), q(t)) \leq \lambda_T + \lambda_S + \epsilon_S(h) + d_{\mathcal{H}\Delta\mathcal{H}}(p(t), q(t))$$

$$\leq \lambda + \hat{e}_S(h) + \delta_S(h) + d_{\mathcal{H}\Delta\mathcal{H}}(p(t), q(t))$$

The first and the fifth inequalities use Lemma A1.  □

Note: The original theorem in Ben David et al., 2006, is slightly different in that their encoder is deterministic and that the decoder is stochastic while ours are in an opposite situation. However, this do not substantially affect the result.

*Appendix A.2. Generalization Bound*

Note: The logarithms in this work are all based on e, i.e., the unit of entropy is nat.

**Lemma A2** (Plug-in Estimation Error [8])**.** *Let $\rho$ be a distribution vector of arbitrary (possible countably infinite) cardinality, and let $\hat{\rho}$ be an empirical estimation of $\rho$ based on a sample of size m. Then, with a probability of at least $1 - \delta$ over the samples,*

$$||\rho - \hat{\rho}||_2 \leq \frac{2 + \sqrt{2log(1/\delta)}}{\sqrt{m}}$$

We apply this bound simultaneously to totally $|\mathcal{Y}| + 2$ terms: $||p(x) - \hat{p}(x)||_2, ||p(y) - \hat{p}(y)||_2, ||p(x|y) - \hat{p}(x|y)||_2$. Therefore, with a probability of at least $1 - \delta$ over the samples, the $|\mathcal{Y}| + 2$ terms above are respectively bounded by $\frac{2 + \sqrt{2log((|\mathcal{Y}|+2)/\delta)}}{\sqrt{m}}$.

**Lemma A3** (Shamir et al. [8])**.** *A real-valued function $\phi$ is defined as follows*

$$\phi(x) = \begin{cases} 0 & x = 0 \\ xlog(1/x) & 0 < x \leq 1/e \\ 1/e & x > 1/e \end{cases} \tag{A1}$$

*$\phi$ is a continuous, monotonically increasing, concave function satisfying $\forall a, b \in [0, 1]$,*

$$|alog(a) - blog(b)| \leq \phi(|a - b|), \tag{A2}$$

**Lemma A4.** *If $X \in [a, b]$, $EX = \mu$, then $VarX \leq (b - \mu)(\mu - a)$.*

**Proof.** (1) If $X \in [0, 1], EX = \mu$, then $VarX = EX^2 - (EX)^2 \leq EX - (EX)^2 = \mu(1 - \mu)$. Equality holds when $EX^2 = EX$, i.e., X = 0 or 1.
(2) Generally, if $X \in [a, b]$, then $\frac{X-a}{b-a} \in [0, 1], E\frac{X-a}{b-a} = \frac{\mu-a}{b-a}$. With the results in (1), we have $Var\frac{X-a}{b-a} \leq \frac{\mu-a}{b-a}(1 - \frac{\mu-a}{b-a})$, i.e., $VarX \leq (b - \mu)(\mu - a)$. Equality holds when $\frac{X-a}{b-a} = 0$ or 1, i.e., X = a or b.  □

**Lemma A5.** *If $x_i \in [0, 1], i = 1, 2, \ldots n, \sum_i^n x_i = 1$, then*

$$\sum_i^n \left( \sqrt{x_i(1 - x_i)} ln(\sqrt{x_i(1 - x_i)}) - \sqrt{x_i} ln x_i \right) \geq 0$$

**Proof.** Let $f(x) = \sqrt{x(1 - x)} ln(\sqrt{x(1 - x)}) - \sqrt{x} ln x$. Let $x_0 \in (0, 1)$ be the zero point of $f(x)$. It is easy to verify that $x_0 \geq 1/2$, $f(x)$ is non-negative and concave on $[0, x_0]$. $\forall x_1, x_2 \in [0, x_0]$, satisfying $x_1 + x_2 \in [0, x_0]$; by the property of concave function, we have

$\frac{f(x_1+x_2)-f(x_1)}{x_2} \leq \frac{f(x_2)}{x_2}, f(x_1+x_2) \leq f(x_1) + f(x_2).$

Now, we prove if $x_i \in [0,1], i = 1, 2, \ldots n, \sum_i^n x_i = 1$, then $f(x_1) + \ldots + f(x_n) \geq 0(*)$.

(1) If $x_i \in [0, 1/2]$, since $f(x)$ is non-negative on $[0, 1/2]$, (*) holds.

(2) If there exists $x_i > 1/2$, since $\sum_i^n x_i = 1$, there is at most one $x_i$ that is larger than $1/2$. We denote it by $x_n$. It is easy to verify that $f(x) + f(1-x) = \sqrt{x}(\sqrt{1-x}-1)lnx + \sqrt{1-x}(\sqrt{x}-1)ln(1-x) \geq 0, x \in [0,1]$. Then, $f(x_1) + \ldots + f(x_n) \geq f(x_1 + x_2) + f(x_3) + \ldots + f(x_n) \geq f(x_1 + .. + x_{n-1}) + f(x_n) = f(x_n) + f(1-x_n) \geq 0$, i.e., (*) holds. □

**Proof of Theorem 3.**

$$
\begin{aligned}
|I(Y;T) - \hat{I}(Y;T)| &\leq |H(T) - \hat{H}(T)| + |H(T|Y) - \hat{H}(T|Y)| \\
&= |H(T) - \hat{H}(T)| + |\sum_y (p(y)H(T|y) - \hat{p}(y)\hat{H}(T|y))| \\
&\leq |H(T) - \hat{H}(T)| + |\sum_y p(y)(H(T|y) - \hat{H}(T|y))| + |\sum_y (p(y) - \hat{p}(y))\hat{H}(T|y)|
\end{aligned}
\tag{A3}
$$

The process of these three parts has some similarities, and we use the first part as an example to illustrate where our proof differs from the original proof. For the first term of (A3),

$$|H(T) - \hat{H}(T)| = |\sum_t p(t)log(p(t)) - \hat{p}(t)log(\hat{p}(t))| \tag{A4}$$

$$\leq \sum_t \phi(|p(t) - \hat{p}(t)|) = \sum_t \phi(|\sum_x p(t|x)(p(x) - \hat{p}(x))|) \tag{A5}$$

$$= \sum_t \phi(|\sum_x (p(t|x) - E_X[p(t|x)])(p(x) - \hat{p}(x))|) \tag{A6}$$

$$= \sum_t \phi(|E_X[(p(t|x) - E_X[p(t|x)])(1 - \frac{\hat{p}(x)}{p(x)})]|) \tag{A7}$$

$$\leq \sum_t \phi\left(\sqrt{Var_X(p(t|x))E_X[(1 - \frac{\hat{p}(x)}{p(x)})^2]}\right) \tag{A8}$$

$$\leq \sum_t \phi\left(\sqrt{Var_X(p(t|x)) \sum_x \frac{1}{min_x p(x)}(p(x) - \hat{p}(x))^2}\right) \tag{A9}$$

$$\leq \sum_t \phi\left(C\sqrt{Var_X(p(t|x))}\right) \tag{A10}$$

where $C = \sqrt{\frac{1}{min_x p(x)}} \frac{2 + \sqrt{2log((|\mathcal{Y}|+2)/\delta)}}{\sqrt{m}}$. (A6) uses $\sum_x p(x) = \sum_x \hat{p}(x) = 1$. (A8) uses Cauchy–Schwarz inequality. (A10) uses Lemma A2. Suppose the number of sample m is large enough subject to $0 < C \leq 1, C\sqrt{(1 - p(t))p(t)} \leq 1/e$.

Our proof begins to deviate from the original proof in (A6), where the original proof add the sample mean and then derive the sample variation: (A5) $= \sum_t \phi(|\sum_x (p(t|x) - \frac{1}{|\mathcal{X}|} \sum_x p(t|x))(p(x) - \hat{p}(x))|) \leq \sum_t \phi\left(\sqrt{|\mathcal{X}|V_x(p(t|x)) \sum_x (p(x) - \hat{p}(x))^2}\right)$. Since here, the two proofs take two totally different ways to process the variance, leading to two different bounds, we treat the variance with our lemmas, while the proof of the original bound uses triangle inequality and an inequality linking KL-divergence and the L1 norm. To grasp the detailed distinction between the two proofs, we recommend reading them. Now, we first continue our proof from (A10) and then show how Shamir et al. [8] processed the variance afterwards.

$$\sum_t \phi\left(C\sqrt{Var_X(p(t|x))}\right) \tag{A11}$$

$$\leq \sum_t \phi\left(C\sqrt{(1-p(t))p(t)}\right) \tag{A12}$$

$$= -\sum_t \left(C\sqrt{(1-p(t))p(t)}\right)log\left(C\sqrt{(1-p(t))p(t)}\right) \tag{A13}$$

$$= -ClogC\sum_t \sqrt{(1-p(t))p(t)} \tag{A14}$$

$$- C\sum_t \sqrt{(1-p(t))p(t)}log(\sqrt{(1-p(t))p(t)}) \tag{A15}$$

$$\leq -\sqrt{(|\mathcal{T}|-1)}ClogC - C\sum_t \sqrt{p(t)}log(p(t)) \tag{A16}$$

$$\leq -\sqrt{(|\mathcal{T}|-1)}ClogC - C\sqrt{\frac{1}{min_t p(t)}}\sum_t p(t)log(p(t)) \tag{A17}$$

$$= C_1\frac{1}{\sqrt{m}}\sqrt{|\mathcal{T}|-1} + C_2\frac{1}{\sqrt{m}}H(T) \tag{A18}$$

where $C_1 = -\sqrt{m}ClogC = \sqrt{\frac{1}{min_x p(x)}}(2 + \sqrt{2log((|\mathcal{Y}|+2)/\delta)})log(1/\sqrt{\frac{1}{min_x p(x)}}\frac{2+\sqrt{2log((|\mathcal{Y}|+2)/\delta)}}{\sqrt{m}})$, $C_2 = \sqrt{m}C\sqrt{\frac{1}{min_t p(t)}} = \sqrt{\frac{1}{min_x p(x)min_t p(t)}}(2 + \sqrt{2log((|\mathcal{Y}|+2)/\delta)})$.

If $p(t) = 0$, then $p(t)logp(t) = 0$. Therefore, there is no need to consider the terms where $p(t) = 0$ in (A17). Therefore, $C_1, C_2$ only depend on $m, \delta, |\mathcal{Y}|, min_x p(x), min_t p(t)$ (here, $min$ refers to the minimum value other than 0). Equation (A12) uses Lemma 1: Since $p(t|x) \in [0,1]$, $E_X[p(t|x)] = p(t)$, then, we have $Var_X(p(t|x)) \leq (1-p(t))p(t)$. Equation (A13) uses the definition of $\phi$ in Lemma A3. Equation (A16) uses Lemma 2 and the Jensen inequality: since $\sqrt{x(1-x)}$ is concave, we have $\sum_t \sqrt{(1-p(t))p(t)} \leq |\mathcal{T}|\sqrt{(1-\frac{1}{|\mathcal{T}|}\sum_t p(t))\frac{1}{|\mathcal{T}|}\sum_t p(t)} = \sqrt{|\mathcal{T}|-1}$.

Here is how Shamir et al. [8] processed the sample variance:

$$\sqrt{V_x(p(t|x))} = p(t)\sqrt{\sum_x(\frac{p(x|t)}{p(x)} - \frac{1}{|\mathcal{X}|}\sum_{x'}\frac{p(x'|t)}{p(x')})^2}$$

$$\leq p(t)[\sqrt{\sum_x(\frac{p(x|t)}{p(x)} - 1)^2} + \sqrt{\sum_x(1 - \frac{1}{|\mathcal{X}|}\sum_{x'}\frac{p(x'|t)}{p(x')})^2}]$$

$$\leq (1 + \frac{1}{\sqrt{|\mathcal{X}|}})||\frac{p(x|t)}{p(x)} - \mathbf{1}||_1 \leq \frac{2}{min_x p(x)}||p(x|t) - p(x)||_1$$

$$\leq \frac{2\sqrt{2log2}}{min_x p(x)}\sqrt{D_{KL}[p(x|t)||p(x)]}$$

For the second term of (A3),

$$\left|\sum_y p(y)(H(T|y) - \hat{H}(T|y))\right| \tag{A19}$$

$$\leq \left|\sum_y p(y)\sum_t(\hat{p}(t|y)log(\hat{p}(t|y)) - p(t|y)log(p(t|y)))\right| \tag{A20}$$

$$\leq \sum_y p(y)\sum_t \phi(|\hat{p}(t|y) - p(t|y)|) \tag{A21}$$

$$= \sum_y p(y)\sum_t \phi\left(\left|\sum_x p(t|x)(\hat{p}(x|y) - p(x|y))\right|\right) \tag{A22}$$

Now use the same technique as that used when processing $|H(T) - \hat{H}(T)|$.

$$\sum_y p(y) \sum_t \phi\left(\left|\sum_x p(t|x)(\hat{p}(x|y) - p(x|y))\right|\right) \tag{A23}$$

$$= \sum_y p(y) \sum_t \phi\left(\left|\sum_x p(x|y)\left(p(t|x) - \sum_x p(t|x)p(x|y)\right)\left(\frac{\hat{p}(x|y)}{p(x|y)} - 1\right)\right|\right) \tag{A24}$$

$$= \sum_y p(y) \sum_t \phi\left(\sqrt{\sum_x p(x|y)(p(t|x) - p(t|y))^2}\sqrt{\sum_x p(x|y)(\frac{\hat{p}(x|y)}{p(x|y)} - 1)^2}\right) \tag{A25}$$

$$\leq \sum_y p(y) \sum_t \phi\left(\sqrt{Var_{X|Y}(p(t|x))\frac{1}{min_x p(x|y)}||\hat{p}(x|y) - p(x|y)||_2^2}\right) \tag{A26}$$

$$\leq \sum_y p(y) \sum_t \phi\left(C\sqrt{p(t|y)(1 - p(t|y))}\right) \tag{A27}$$

$$= -\sum_y p(y) \sum_t C\sqrt{p(t|y)(1 - p(t|y))}log\left(C\sqrt{p(t|y)(1 - p(t|y))}\right) \tag{A28}$$

$$= -\sum_y p(y) \sum_t ClogC\sqrt{p(t|y)(1 - p(t|y))} - \sum_y p(y) \sum_t C\sqrt{p(t|y)(1 - p(t|y))}$$
$$log\left(\sqrt{p(t|y)(1 - p(t|y))}\right) \tag{A29}$$

$$\leq -ClogC\sqrt{|\mathcal{T}| - 1} + C\sqrt{\frac{1}{min_{t,y}p(t|y)}}H(T|Y) \tag{A30}$$

$$= C_3\frac{1}{\sqrt{m}}\sqrt{|\mathcal{T}| - 1} + C_4\frac{1}{\sqrt{m}}H(T|Y) \tag{A31}$$

where $C = \sqrt{\frac{1}{min_{x,y}p(x|y)}}\frac{2 + \sqrt{2log((|\mathcal{Y}|+2)/\delta)}}{\sqrt{m}}$. Suppose that the number of sample m is large enough subject to $0 < C \leq 1$, $C\sqrt{p(t|y)(1 - p(t|y))} \leq 1/e$. $C_3 = -\sqrt{m}ClogC = \sqrt{\frac{1}{min_{x,y}p(x|y)}}$ $(2 + \sqrt{\frac{2log((|\mathcal{Y}|+2)}{\delta}})log\left(1/\sqrt{\frac{1}{min_{x,y}p(x|y)}}\frac{2 + \sqrt{2log((|\mathcal{Y}|+2)/\delta)}}{\sqrt{m}}\right)$, $C_4 = C\sqrt{\frac{1}{min_{t,y}p(t|y)}}$ $= \sqrt{\frac{1}{min_{x,y}p(x|y)min_{t,y}p(t|y)}}(2 + \sqrt{2log((|\mathcal{Y}|+2)/\delta)})$ only depending on $m$, $\delta$, $|\mathcal{Y}|$, $min_{x,y}p(x|y)$, $min_{t,y}p(t|y)$ (here, *min* refers to the minimum value other than 0).

With the same technique as above, the third term of (A3) can be bounded as follows:

$$\left|\sum_y (p(y) - \hat{p}(y))\hat{H}(T|y)\right| \tag{A32}$$

$$= \left|\sum_y \hat{p}(y)(\frac{p(y)}{\hat{p}(y)} - 1)(\hat{H}(T|y) - \sum_y \hat{p}(y)\hat{H}(T|y)\right| \tag{A33}$$

$$\leq ||p(y) - \hat{p}(y)||\sqrt{\frac{1}{min_y \hat{p}(y)}Var_{\hat{Y}}(\hat{H}(T|y))} \tag{A34}$$

$$= C_5\frac{1}{\sqrt{m}}\sqrt{Var_{\hat{Y}}(\hat{H}(T|y))} \tag{A35}$$

$$\leq C_5\frac{1}{\sqrt{m}}\sqrt{(log|\mathcal{T}| - \hat{H}(T|Y))\hat{H}(T|Y)} \tag{A36}$$

where $C_5 = \sqrt{\frac{1}{min_y \hat{p}(y)}}(2 + \sqrt{2log((|\mathcal{Y}| + 2)/\delta)})$.

From (A18), (A31), and (A36), we conclude Theorem 3:

$$|I(Y;T) - \hat{I}(Y;T)| \leq$$
$$\frac{1}{\sqrt{m}}\left((C_1 + C_3)\sqrt{|\mathcal{T}| - 1} + C_2 H(T) + C_4 H(T|Y) + C_5\sqrt{(log|\mathcal{T}| - \hat{H}(T|Y))\hat{H}(T|Y)}\right)$$

□

Here we provide some discussions of the results.

A. A comparison of the order of the previous bound and our bound.

$$Bound_{Previous} \sim -4\sqrt{2log(2)}D|\mathcal{X}|log(2\sqrt{2log(2)}D|\mathcal{X}|)\sqrt{|\mathcal{T}|}\sqrt{I(X;T)} +$$
$$4\sqrt{2log(2)}D|\mathcal{X}||\mathcal{T}|^{3/4}I(X;T)^{1/4} + 2D|\mathcal{Y}|I(X;T)$$

$$Bound_{Ours} \sim -2D\sqrt{|\mathcal{X}|}log(D\sqrt{|\mathcal{X}|})\sqrt{|\mathcal{T}|} + D\sqrt{|\mathcal{X}|}\sqrt{|\mathcal{T}|}(H(T|Y) + H(T|Y)) +$$
$$D\sqrt{|\mathcal{Y}|}\sqrt{(log|\mathcal{T}| - H(T|Y))H(T|Y)}$$

Some explanations

- $D = \frac{2+\sqrt{2log((|\mathcal{Y}|+2)/\delta)}}{\sqrt{m}}$ is a constant.
- The sample size $m$ should be large enough such that $log(D\sqrt{|\mathcal{X}|}) < 0$ and $log(2\sqrt{2log(2)}D|\mathcal{X}|) < 0$.
- Since $|\mathcal{X}|$, $|\mathcal{T}|$, $|\mathcal{Y}|$ can be very large in real-world data, the magnitude of the two bounds is mainly controlled by them.
- $I(X;T)$, $H(T|Y)$, $H(T)$ are bounded by $log|\mathcal{T}|$. Their effects to the magnitude of the bounds are small. They can be viewed as the same order of $log|\mathcal{T}|$.

Finally, the conclusion is that the IB bound is $\sqrt{|\mathcal{X}|}$ times larger than the DIB bound for the first two terms and $\sqrt{|\mathcal{Y}|}$ times larger for the third term.

B. Constraints for sample size

$$0 < \sqrt{\frac{1}{min_x p(x)}}\frac{2+\sqrt{2log((|\mathcal{Y}|+2)/\delta)}}{\sqrt{m}} \leq 1 \tag{A37}$$

$$\sqrt{\frac{1}{min_x p(x)}}\frac{2+\sqrt{2log((|\mathcal{Y}|+2)/\delta)}}{\sqrt{m}}\sqrt{(1-p(t))p(t)} \leq 1/e \tag{A38}$$

$$0 < \sqrt{\frac{1}{min_x p(x|y)}}\frac{2+\sqrt{2log((|\mathcal{Y}|+2)/\delta)}}{\sqrt{m}} \leq 1 \tag{A39}$$

$$\sqrt{\frac{1}{min_x p(x|y)}}\frac{2+\sqrt{2log((|\mathcal{Y}|+2)/\delta)}}{\sqrt{m}}\sqrt{p(t|y)(1-p(t|y))} \leq 1/e \tag{A40}$$

The IB bound in Shamir et al. [8] has similar constraints.

$$0 < \sqrt{\frac{2+\sqrt{2log((|\mathcal{Y}|+2)/\delta)}}{m}}\frac{2\sqrt{2log(2)}}{min_x p(x)} < 1 \tag{A41}$$

$$\sqrt{\frac{2+\sqrt{2log((|\mathcal{Y}|+2)/\delta)}}{m}}\frac{2\sqrt{2log(2)}}{min_x p(x)}$$
$$p(t)\sqrt{D_{KL}[p(x|t)||p(x)]} < 1/e \tag{A42}$$

They discussed how IB bound trivially holds when (A42) is not satisfied, but neglecting the constraint (A41). Our simulations in Section 4.1.2 compare the sample size the two bounds needs.

C. Why does DIB have a better generalization than IB? The main conclusion in Section 3.1 is explained as follows.

1. The SG bound in Theorem 3 is controlled by $H(T)$ and $H(T|Y)$, which suggests that minimizing $H(T)$ and $H(T|Y)$ narrows the source generalization gap.
2. $H(T)$ and $H(T|Y)$ are equivalently DIB regularizers (see Section 3.2).
3. $I(X;T)$ and $I(X;T|Y)$ are equivalently IB regularizers. This can be proved in the same way as (2): As is assumed in IB, $Y - X - T$ is a Markov chain, so $H(T|X,Y) = H(T|X)$.

$$I(X;T) = H(T) - H(T|X) = I(Y;T) + H(T|Y) - H(T|X)$$
$$= I(Y;T) + H(T|Y) - H(T|X,Y) = I(Y;T) + I(X;T|Y)$$

$L_{IB} = I(X;T) - \beta I(Y;T) = I(X;T|Y) - (\beta - 1)I(Y;T), \beta > 1$

Therefore, $I(X;T|Y)$ is an equivalently IB regularizer, which is also known as CEB (Fischer 2020).

4. DIB achieves lower $H(T)$ and $H(T|Y)$ than IB.
$H(T) = I(X;T) + H(T|X), H(T|Y) = I(X;T|Y) + H(T|X)$. DIB minimizes $H(T)$ and $H(T|Y)$, while IB minimizes part of them according to (2) and (3). Specifically, DIB's optimal $p(t|x)$ is deterministic, i.e., $0 = H(T|X)_{DIB} \leq H(T|X)_{IB}$, while $I(X;T)_{DIB} \approx I(X;T)_{IB}$ (Strouse and Schwab, 2017b). Therefore, DIB's optimal solution has smaller $H(T)$ and $H(T|Y)$ than that of IB, as is found in the simulations (Figure A3 in the Appendix C and Figure 2 in [7]).
5. Finally according to (1) and (4), DIB achieves lower SG than IB in the sense of Theorem 3.

*Appendix A.3. Representation Discrepancy*

**Proof of Proposition 1.**

$$d_{\mathcal{H}\Delta\mathcal{H}}(p(t), q(t)) = sup_{h \in \mathcal{H}} |E_{t \sim p(t)} I_{\{h^*(t) \neq h(t)\}} - E_{t \sim q(t)} I_{\{h^*(t) \neq h(t)\}}| \tag{A43}$$

$$\leq sup_{h \in \mathcal{H}} \sum_{\{t|h^*(t) \neq h(t)\}} |p(t) - q(t)| \leq sup_{h \in \mathcal{H}} \sum_{\{t|h^*(t) \neq h(t)\}} |\hat{p}(t) - \hat{q}(t)| + \epsilon \tag{A44}$$

$$= sup_{h \in \mathcal{H}} \sum_{\{t|h^*(t) \neq h(t)\}} |\frac{1}{m} \sum_{x \in \mathcal{X}_S} p(t|x) - \frac{1}{m} \sum_{x \in \mathcal{X}_T} p(t|x)| + \epsilon \tag{A45}$$

$$\leq \frac{1}{m} \sum_{(x_S, x_T) \in \mathcal{CP}} sup_{h \in \mathcal{H}} \sum_{\{t|h^*(t) \neq h(t)\}} |p(t|x_S) - p(t|x_T)| + \epsilon \tag{A46}$$

$$\leq \frac{1}{m} \sum_{(x_S, x_T) \in \mathcal{CP}} \|p(t|x_S) - p(t|x_T)\|_1 + \epsilon \tag{A47}$$

Equation (A44) uses the Cauchy–Schwarz inequality and Lemma A2. $\epsilon$ is small when the sample size is large. □

**Proof of** (22).

$$\|p(t|x_S) - p(t|x_T)\|_1 \tag{A48}$$

$$= \Pi_{i=1}^d \int_R \left| \frac{1}{\sqrt{2\pi}\sigma_i} \left( e^{-\frac{(x_i - \mu_{1i})^2}{2\sigma_i^2}} - e^{-\frac{(x_i - \mu_{2i})^2}{2\sigma_i^2}} \right) \right| dx_i \tag{A49}$$

$$= \Pi_{i=1}^d \int_R \left| \frac{1}{\sqrt{2\pi}\sigma_i} \left( e^{-\frac{x_i^2}{2\sigma_i^2}} - e^{-\frac{(x_i - \mu_{2i} + \mu_{1i})^2}{2\sigma_i^2}} \right) \right| dx_i \tag{A50}$$

$$= \Pi_{i=1}^d 2 \left( \int_{-\infty}^{\frac{|\mu_{1i} - \mu_{2i}|}{2\sigma_i}} \frac{1}{\sqrt{2\pi}} e^{-\frac{x_i^2}{2}} - \int_{\frac{|\mu_{1i} - \mu_{2i}|}{2\sigma_i}}^{\infty} \frac{1}{\sqrt{2\pi}} e^{-\frac{x_i^2}{2}} \right) dx_i \tag{A51}$$

$$= \Pi_{i=1}^d \left( 4 \int_{-\infty}^{\frac{|\mu_{1i} - \mu_{2i}|}{2\sigma_i}} \frac{1}{\sqrt{2\pi}} e^{-\frac{x_i^2}{2}} dx_i - 2 \right) \tag{A52}$$

$$= \Pi_{i=1}^d \left( 4\Phi \left( \frac{|\mu_{1i} - \mu_{2i}|}{2\sigma_i} \right) - 2 \right) \tag{A53}$$

□

In order to simplify the calculation, we assumed that the source and target domains have the same sample sizes and a pair-wise correspondence, but in fact, the "pair-wise" assumption is not essential. What is essential is the intrinsic correspondence between the two domains. For example, all the zeros in MNIST digits and all the zeros in USPS digits have a semantic correspondence. This is a reasonable assumption for transfer learning because when the source domain and target domain are not related to each other, brute-force transfer may be unsuccessful (Pan et al., 2009).

When the sample size on two domains are different and the "pair-wise" correspondence does not hold, we can take the correlated instances in two domains as a general form of correspondent pairs, named correspondent group pairs. Specifically, the instances are partitioned into, say, n parts: $\mathcal{X}_S = \bigsqcup_i^n G_{S,i}$, $\mathcal{X}_T = \bigsqcup_i^n G_{T,i}$, where $\bigsqcup$ is a disjoint union. $\forall$ fixed i, the instances in $G_{S,i}$ and the instances in $G_{T,i}$ are correlated. The collection of all the correspondent group pairs is denoted by $\mathcal{CP} \triangleq \{(G_{S,i}, G_{T,i})_{i=1...n}\}$. Take digit recognition as an example; the instances in two domains can both be partitioned into n = 10 groups. $G_{S,i}$ and $G_{T,i}$ contain the instances of digit $(i-1)$. Then, $(G_{S,i}, G_{T,i})$ is a correspondent group pair.

The comparison of IB and DIB on RD can be explicitly proven under the assumption of diagonal Gaussian and assumption of balanced samples ($\forall i, j \in \{1 \ldots n\}, |G_{S,i}| = |G_{S,j}|, |G_{T,i}| = |G_{T,j}|$, where $|\cdot|$ denotes the cardinality of a set, i.e., $|\mathcal{X}_S| = n|G_{S,i}|, |\mathcal{X}_T| = n|G_{T,i}|$). This proof is a generalized form of the proof of Proposition 1.

$$(A46) = sup_{h \in \mathcal{H}} \sum_{\{t|h^*(t) \neq h(t)\}} |\hat{p}(t) - \hat{q}(t)| + \epsilon \tag{A54}$$

$$= sup_{h \in \mathcal{H}} \sum_{\{t|h^*(t) \neq h(t)\}} \left| \frac{1}{|\mathcal{X}_S|} \sum_{x_S \in \mathcal{X}_S} p(t|x_S) - \frac{1}{|\mathcal{X}_T|} \sum_{x_T \in \mathcal{X}_T} p(t|x_T) \right| + \epsilon \tag{A55}$$

$$= sup_{h \in \mathcal{H}} \sum_{\{t|h^*(t) \neq h(t)\}} \frac{1}{n} \sum_{i=1}^n \left| \frac{1}{|G_{S,i}|} \sum_{x_S \in G_{S,i}} p(t|x_S) - \frac{1}{|G_{T,i}|} \sum_{x_T \in G_{T,i}} p(t|x_T) \right| + \epsilon \tag{A56}$$

$$= sup_{h \in \mathcal{H}} \sum_{\{t|h^*(t) \neq h(t)\}} \frac{1}{n} \sum_{i=1}^n \left| \frac{1}{|G_{S,i}| \cdot |G_{T,i}|} \sum_{x_S \in G_{S,i}, x_T \in G_{T,i}} \left( p(t|x_S) - p(t|x_T) \right) \right| + \epsilon \tag{A57}$$

$$\leq \frac{1}{n} \sum_{i=1}^n \frac{1}{|G_{S,i}| \cdot |G_{T,i}|} \sum_{x_S \in G_{S,i}, x_T \in G_{T,i}} \|p(t|x_S) - p(t|x_T)\|_1 + \epsilon \tag{A58}$$

Then, with Equation (22) and the analysis in our paper, IB's RD is smaller than that of DIB.

**Appendix B. Iterative and Variational Algorithm for EIB**

Please note that Strouse and Schwab [7] constructed a generalized IB expression that has the same form as EIB. They used a generalized IB to solve the iterative algorithm of DIB and to raise it as an potential method for soft clustering, but did not study it in depth, let alone use it in transfer learning. In this section, we provide a traditional iterative algorithm of EIB and a proof of convergence, the proof of which is quite similar to that of IB. After that, we give a variational algorithm for EIB, which is more useful in practice.

*Appendix B.1. An Iterative Algorithm of EIB and a Proof of Convergence*

**Proposition A1.** *The first-order variation of $L_{EIB}$ at $p(t|x)$ along $h(t|x)$ is as follows:*

$$\delta L[p(t|x)] = \sum_{x,t} p(x)h(t|x) \log \frac{p(t|x)^\alpha}{p(t)} + (\alpha - 1)\sum_{x,t} h(t|x)p(x) - \beta \sum_{x,y,t} p(x,y)h(t|x) \log \frac{p(t|y)}{p(t)} \quad (A59)$$

Let the first-order variation equal 0; the expression of the optimal solution $p(t|x)$ can then be obtained:

$$p(t|x) = \frac{1}{Z(x,\alpha,\beta)} exp(\frac{1}{\alpha}(log\,p(t) - \beta D_{KL}[p(y|x)||p(y|t)])) \quad (A60)$$

where $Z(x,\alpha,\beta)$ is a normalization coefficient. Now, we obtain the iterative algorithm for solving EIB in Algorithm A1.

---

**Algorithm A1** The elastic information bottleneck method.

---

**input:** $p(x,y)$: Joint probability distribution;
1: $\alpha,\beta$: Parameters
2: $|\mathcal{T}|$: Cardinality of set T/maximum number of categories*;
3: Initialization: $p^{(0)}(t|x)$;n=0;
4: **repeat**
5:      $p^{(n)}(t) = \sum_x p(x)p^{(n)}(t|x), \forall t \in \mathcal{T}$;
6:      $p^{(n)}(y|t) = \frac{\sum_x p^{(n)}(t|x)p(x,y)}{p^{(n)}(t)}, \forall t \in \mathcal{T}, \forall y \in \mathcal{Y}$;
7:      $p^{(n+1)}(t|x) = \frac{1}{Z(x,\alpha,\beta)} exp(\frac{1}{\alpha}(log\,p^{(n)}(t) - \beta D_{KL}[p(y|x)||p^{(n)}(y|t)]))$;
8:      n=n+1;
9: **until** $p(t|x)$ converge.
**output:** Optimal solution $p(t|x)$;

---

**Comments.** *Note that the actual number of representation clusters can be smaller than $|\mathcal{T}|$ because EIB minimizes H(T), resulting in $p(t) = 0$ for some t.*

**Proposition A2** (Convergence of EIB algorithm). *For the three iterative formulas of EIB, $L_{EIB}$ will decrease every time an iterative formula is updated. Given that $\alpha$, $\beta$, $L_{EIB}$ have lower bounds, the EIB algorithm converges.*

**Proof.** Let $F_{EIB}$ be the negative logarithmic expectation of the normalized coefficient, which is called the free energy of the system in physics.

$$F_{EIB} = -\sum_{x,t} p(x)p(t|x)log\,Z(x,\beta,\alpha)$$

□

**Lemma A6.** $F_{EIB}$ *is non-negative. If any two terms of* $p(t|x), p(t), p(t|y)$ *are fixed,* $F_{EIB}$ *is convex with respect to the third term.*

**Proof.**

$$
F_{EIB} = \sum_{x,t} p(x)p(t|x)log\frac{p(t|x)^{\alpha}}{p(t)} + \beta \sum_{x,t,y} p(t|x)p(y,x)log\frac{p(y|x)}{p(y|t)} = \sum_{x} \alpha D_{KL}[p(t|x)||p(t)] \tag{A61}
$$
$$
+ (1-\alpha) \sum_{x,t} p(x)p(t|x)log\frac{1}{p(t)} + \beta \sum_{x,t} p(x)p(t|x)D_{KL}[p(y|x)||p(y|t)]
$$

Because KL-divergence is non-negative, $F_{EIB}$ is non-negative. Since the function $g_1(x) = xlnx, 0 < x < 1$ is strictly convex with respect to x, and the function $g_2(x) = -lnx, 0 < x < 1$ is strictly convex with respect to x; obviously, $F_{EIB}$ is strictly convex with respect to $p(t|x), p(t), p(t|y)$. □

**Lemma A7.** *When* $p(t|x), p(t), p(t|y)$ *is updated according to the EIB iteration formula,* $F_{EIB}$ *will not increase.*

**Proof.** First, we prove that when $p(t|x)$ is updated, $F_{EIB}$ will not increase. Let us add the normalization condition, $\hat{F}_{EIB} \triangleq F_{EIB} + \sum_x \lambda(x)[\sum_t p(t|x) - 1]$, and let the derivative of $\hat{F}_{EIB}$ with respect to $p(t|x)$ be 0:

$$
p(x)log\frac{p(t|x)^{\alpha}}{p(t)} + \alpha p(x) + \beta \sum_y p(y,x)log\frac{p(y|x)}{p(y|t)} + \lambda(x) = 0
$$
$$
log\frac{p(t|x)^{\alpha}}{p(t)} + \beta \sum_y p(y|x)log\frac{p(y|x)}{p(y|t)} + \hat{\lambda}(x) = 0 \tag{A62}
$$
$$
p(t|x) = \frac{1}{Z(x,\alpha,\beta)}exp(\frac{logp(t) - \beta D_{KL}[p(y|x)||p(y|t)]}{\alpha})
$$

The above formula happens to be the iterative formula of $p(t|x)$. Since $\hat{F}_{EIB}$ is strictly convex, when $p(t|x)$ is updated, $F_{EIB}$ will not increase.

Similarly, it is easy to verify that the derivative of $F_{EIB} + \lambda[\sum_t p(t) - 1]$ with respect to $p(t)$ being 0 is exactly the iterative formula of $p(t)$ and that the derivative of $F_{EIB} + \sum_t \lambda(t)[\sum_y p(y|t) - 1]$ with respect to $p(y|t)$ being 0 is exactly the iterative formula of $p(y|t)$.

According to the above two lemmas, every iteration update makes $F_{EIB}$ non-increasing. Since $F_{EIB}$ has a lower bound, the iterative process converges. □

*Appendix B.2. Variational EIB*

VIB [15] is a classical variational approach for IB, but its assumption that p(t) is standard Gaussian distribution contradicts minimizing H(T) in DIB. Therefore, we utilize the variational approaches similar to CEB [2] for variantional EIB. Figure A1 illustrates the network structure.

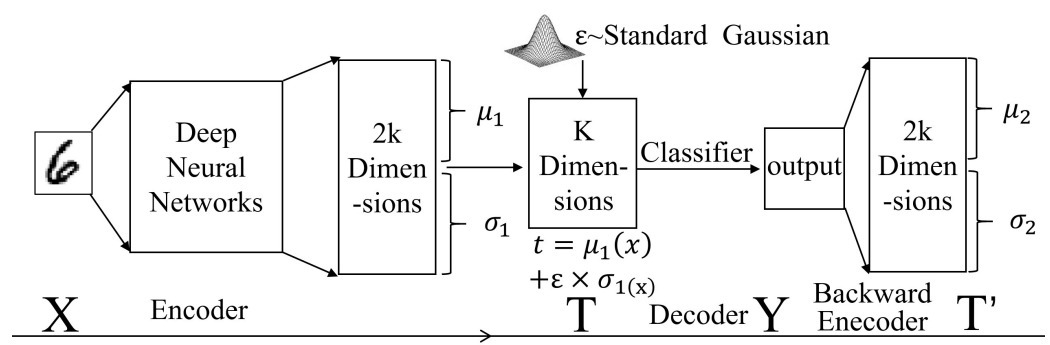

**Figure A1.** The network structure of variantional EIB.

The loss function of variantional EIB is derived as follows:

$$L_{EIB} = H(T|Y) - \alpha H(T|X) - \beta I(Y,T) \tag{A63}$$

$$= \int p(t|x)p(x,y)\left(log\frac{p^\alpha(t|x)}{p(t|y)} - \beta log p(y|t)\right)\mathrm{d}x\mathrm{d}t\mathrm{d}y - H(Y) \tag{A64}$$

$$\leq \int p(t|x)p(x,y)\left(log\frac{p^\alpha(t|x)}{b(t|y)} - \beta log q(y|t)\right)\mathrm{d}x\mathrm{d}t\mathrm{d}y - H(Y) \tag{A65}$$

where the decoder $q(y|t)$ is a variantional approximation of $p(y|t)$ and the backward encoder $b(t|y)$ is a variantional approximation of $p(t|y)$. The encoder outputs the K-dimensional expectations $\boldsymbol{\mu_1}$ and K-dimensional diagonal elements of covariance matrix $\boldsymbol{\sigma_1}$. Similarly, the backward encoder outputs $\boldsymbol{\mu_2}$ and $\boldsymbol{\sigma_2}$. Then, the regularization term can be written as follows:

$$\int p(t|x)log\frac{p^\alpha(t|x)}{b(t|y)}\mathrm{d}t \tag{A66}$$

$$= \prod_{i=1}^{K}\int_{\mathbb{R}}\frac{1}{\sqrt{2\pi}\sigma_{1,i}}exp(-\frac{(x_i-\mu_{1,i})^2}{2\sigma_{1,i}^2})ln(\frac{(\prod_{j=1}^{K}\frac{1}{\sqrt{2\pi}\sigma_{1,j}}exp(-\frac{(x_j-\mu_{1,j})^2}{2\sigma_{1,j}^2}))^\alpha}{\prod_{j=1}^{K}\frac{1}{\sqrt{2\pi}\sigma_{2,j}}exp(-\frac{(x_j-\mu_{2,j})^2}{2\sigma_{2,j}^2})})\mathrm{d}x_i \tag{A67}$$

$$= \sum_{j=1}^{K}\left[\frac{(1-\alpha)}{2}ln(2\pi) - \alpha ln\sigma_{1,j} + ln\sigma_{2,j} - \frac{\alpha}{2} + \frac{\mu_{1,j}^2+\mu_{2,j}^2-2\mu_{1,j}\mu_{2,j}+\sigma_{1,j}^2}{2\sigma_{2,j}^2}\right] \tag{A68}$$

Using the empirical distribution to approximate the joint distribution p(x,y), the loss function of variational EIB is as follows:

$$L = \frac{1}{m}\sum_{n=1}^{m}\int p(t|x_n)\left(log\frac{p^\alpha(t|x_n)}{p(t|y_n)} - \beta log q(y_n|t)\right)\mathrm{d}t = \beta CE[softmax(y_n)||q(y_n|t_n)] + \tag{A69}$$

$$\frac{1}{m}\sum_{n=1}^{m}\sum_{j=1}^{K}\left[\frac{\mu_{1,j,n}^2+\mu_{2,j,n}^2-2\mu_{1,j,n}\mu_{2,j,n}+\sigma_{1,j,n}^2}{2\sigma_{2,j,n}^2} + \frac{(1-\alpha)}{2}ln(2\pi) - \alpha ln\sigma_{1,j,n} + ln\sigma_{2,j,n} - \frac{\alpha}{2}\right] \tag{A70}$$

## Appendix C. Additional Experiments: EIB on Information Plane

We calculate the optimal solutions of EIB by the iterative algorithm and plot them on the information planes. Figure A2 reveals that when $\alpha$ decreases or $\beta$ increases, I(Y;T) and I(X;T) grow while $H(T|X)$ drops. Figure A3 shows that $\alpha$ and $\beta$ can precisely adjust the position of the optimal solution on the information plane, which is closely related to the clustering property.

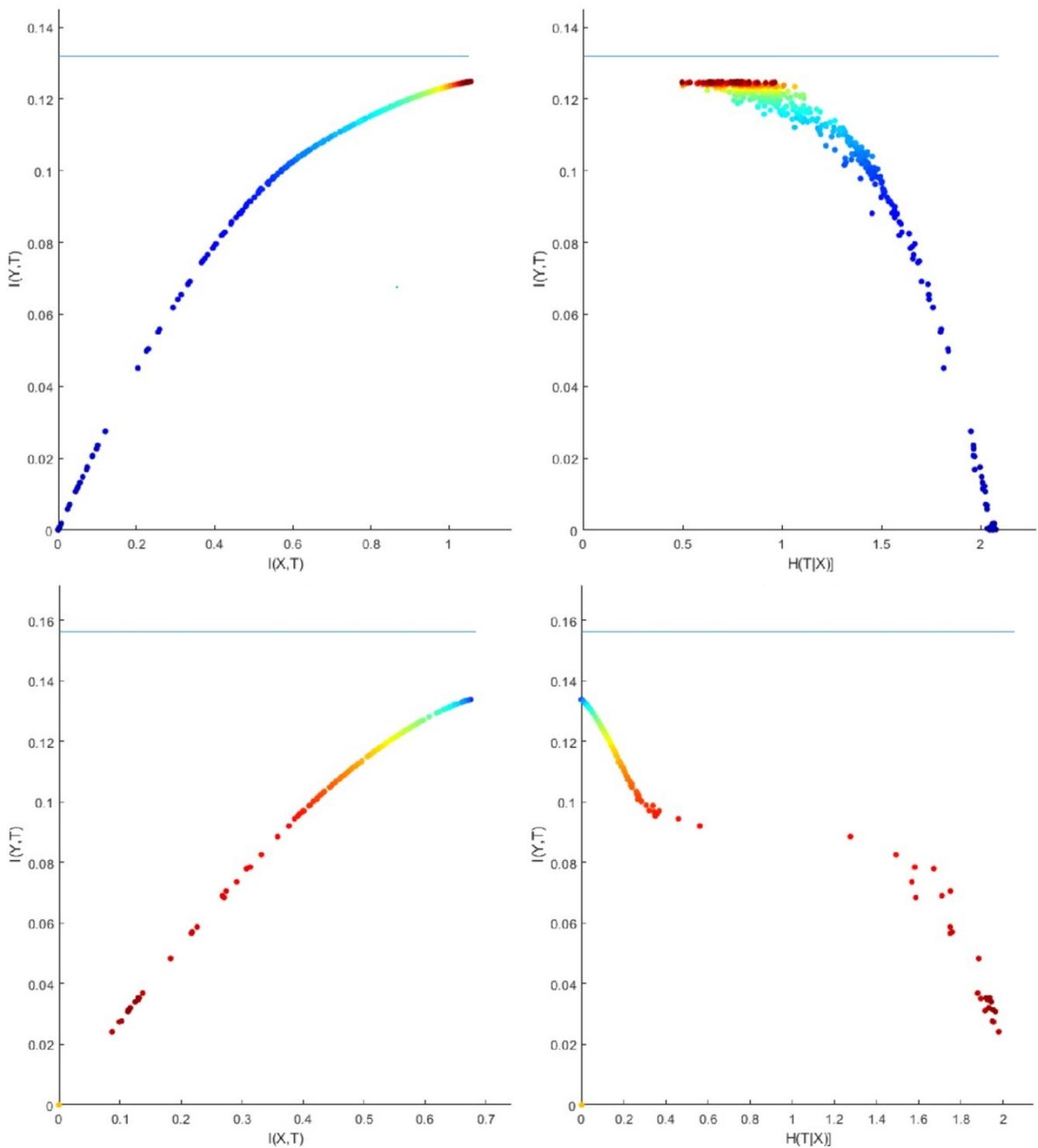

**Figure A2. Top:** $\alpha = 1$; $\beta$ takes 500 points evenly in [1.5, 51.5]; $\beta$ from small to large corresponds to the colors started from blue to red. **Bottom:** $\beta = 4.5$; $\alpha$ takes 201 points evenly in [0, 1]; $\alpha$ from small to large corresponds to the colors starting from blue to red. The blue horizontal line is H(Y), which is the upper bound of I(Y,T).

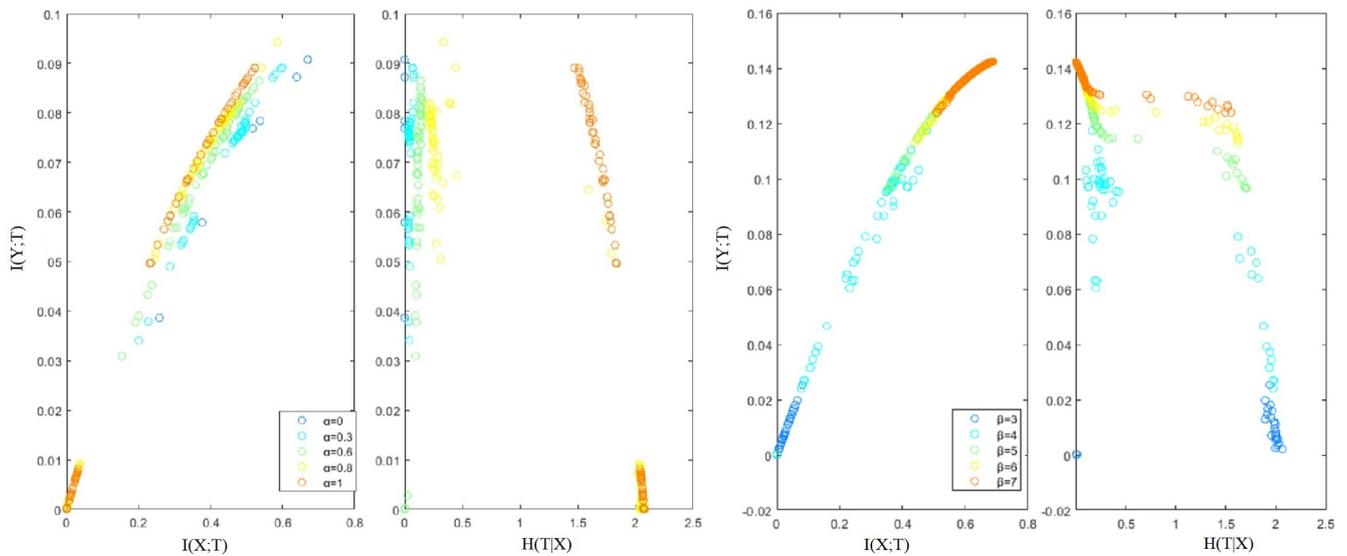

**Figure A3. Left:** Each curve is drawn by altering $\beta$. **Right:** Each curve is drawn by altering $\alpha$.

## Appendix D. Experimental Details

Our codes of variational EIB are based on the code of VIB [15].

For variational EIB on toy data, the dimension of Gaussian representations is $K = 5$. The model was trained with 50 epochs, a 100 batch size, and a 0.01 initial learning rate, and the experiments were repeated with different random seeds.

For variational EIB on MNIST, the dimension of Gaussian representations is $K = 256$. We trained the variational EIB model, with 50 training epochs, a 100 batch size, and a 1e-4 initial learning rate, and the experiments were repeated with different random seeds.

Our codes of EIB-DFA-MCD are based on the code of DFA-MCD Wang et al. [33]. The pseudocode of EIB-DFA-MCD is shown in Algorithm A2:

---

**Algorithm A2** EIB-DFA-MCD.

---

**Input:** Source samples:$(X_s, Y_s)$; Target samples:$(X_t)$; Parameters:$\alpha, \beta, \lambda$.
1: **for** epoch=1 to epoch-number **do**
2:     **for** batch=1 to batch-number **do**
3:         Step 1: Update $E, D_1, D_2, BE_1, BE_2$ to minimize
4:         $Loss_1 = Loss_{EIB}(E, D_1, BE_1) + Loss_{EIB}(E, D_2, BE_2)$;
5:         Step 2: Update $D_1, D_2, BE_1, BE_2$ to minimize
6:         $Loss_2 = Loss_{EIB}(E, D_1) + Loss_{EIB}(E, D_2) - KL[\hat{Y}_{t1}||\hat{Y}_{t2}]$;
7:         Step 3: Update $E$ to minimize $Loss_3 = KL[\hat{Y}_{t1}||\hat{Y}_{t2}] + \lambda KL[\hat{X}_s||\hat{X}_t]$ ;
8:     **end for**
9: **end for**
**Output:** $E, D_1, D_2$.

---

where $E$ stands for the encoder, $D_1$ and $D_2$ stand for the decoders, $BE_1$ and $BE_2$ stand for the backward encoders, $\hat{Y}_{t1}$ and $\hat{Y}_{t2}$ stand for the target label predictions by the two decoders, and $\hat{X}_s$ and $\hat{X}_t$ stand for the target and source reconstructed instances. The dimension of Gaussian representations is $K = 768$. The model was trained with 200 epochs, a 128 batch size, and a 0.0002 initial learning rate, and the experiments were repeated with different random seeds. An OUr computing infrastructure was used to run the experiments: GPU model; memory: 24,268 MiB; operating system: linux; and Pytorch: 1.9.0+cu111.

## References

1. Tishby, N.; Pereira, F.C.; Bialek, W. The information bottleneck method. *arXiv* **2000**, arXiv:physics/0004057.
2. Fischer, I. The conditional entropy bottleneck. *Entropy* **2020**, *22*, 999. [CrossRef] [PubMed]
3. Chechik, G.; Globerson, A.; Tishby, N.; Weiss, Y. Information bottleneck for Gaussian variables. *J. Mach. Learn. Res.* **2005**, *6*, 165–188.
4. Slonim, N.; Friedman, N.; Tishby, N. Multivariate Information Bottleneck. *Neural Comput.* **2006**, *18*, 1739–1789. [CrossRef]
5. Aguerri, I.E.; Zaidi, A. Distributed variational representation learning. *IEEE Trans. Pattern Anal. Mach. Intell.* **2019**, *43*, 120–138. [CrossRef] [PubMed]
6. Kolchinsky, A.; Tracey, B.D.; Van Kuyk, S. Caveats for information bottleneck in deterministic scenarios. In Proceedings of the International Conference on Learning Representations (ICLR), New Orleans, LA, USA, 6–9 May 2019.
7. Strouse, D.; Schwab, D.J. The deterministic information bottleneck. *Neural Comput.* **2017**, *29*, 1611–1630. [CrossRef]
8. Shamir, O.; Sabato, S.; Tishby, N. Learning and generalization with the information bottleneck. *Theor. Comput. Sci.* **2010**, *411*, 2696–2711. [CrossRef]
9. Wang, B.; Wang, S.; Cheng, Y.; Gan, Z.; Jia, R.; Li, B.; Liu, J. Infobert: Improving robustness of language models from an information theoretic perspective. *arXiv* **2020**, arXiv:2010.02329.
10. Shwartz-Ziv, R.; Tishby, N. Opening the black box of deep neural networks via information. *arXiv* **2017**, arXiv:1703.00810.
11. Tishby, N.; Zaslavsky, N. Deep learning and the information bottleneck principle. In Proceedings of the 2015 IEEE Information Theory Workshop (ITW), Jeju Island, Korea, 11–15 October 2015; pp. 1–5.
12. Saxe, A.M.; Bansal, Y.; Dapello, J.; Advani, M.; Kolchinsky, A.; Tracey, B.D.; Cox, D.D. On the information bottleneck theory of deep learning. *J. Stat. Mech. Theory Exp.* **2019**, *2019*, 124020. [CrossRef]
13. Slonim, N. The Information Bottleneck: Theory and Applications. Ph.D. Thesis, Hebrew University of Jerusalem, Jerusalem, Israel, 2002.
14. Slonim, N.; Tishby, N. Agglomerative information bottleneck. *Adv. Neural Inf. Process. Syst.* **1999**, *12*, 617–623.
15. Alemi, A.A.; Fischer, I.; Dillon, J.V.; Murphy, K. Deep variational information bottleneck. *arXiv* **2016**, arXiv:1612.00410. Available online: https://github.com/1Konny/VIB-pytorch (accessed on 21 April 2012).
16. Higgins, I.; Matthey, L.; Pal, A.; Burgess, C.; Glorot, X.; Botvinick, M.; Mohamed, S.; Lerchner, A. beta-vae: Learning basic visual concepts with a constrained variational framework. In Proceedings of the 5th International Conference on Learning Representations (ICLR 2017), Toulon, France, 24–26 April 2017.
17. Wu, T.; Ren, H.; Li, P.; Leskovec, J. Graph Information Bottleneck. *Adv. Neural Inf. Process. Syst.* **2020**, *33*, 20437–20448.
18. Achille, A.; Soatto, S. Information dropout: Learning optimal representations through noisy computation. *IEEE Trans. Pattern Anal. Mach. Intell.* **2018**, *40*, 2897–2905. [CrossRef]
19. Dubois, Y.; Kiela, D.; Schwab, D.J.; Vedantam, R. Learning optimal representations with the decodable information bottleneck. *Adv. Neural Inf. Process. Syst.* **2020**, *33*, 18674–18690.
20. Wang, Z.; Huang, S.L.; Kuruoglu, E.E.; Sun, J.; Chen, X.; Zheng, Y. PAC-Bayes Information Bottleneck. *arXiv* **2021**, arXiv:2109.14509.
21. Strouse, D.; Schwab, D.J. The information bottleneck and geometric clustering. *arXiv* **2017**, arXiv:1712.09657.
22. Goldfeld, Z.; Polyanskiy, Y. The information bottleneck problem and its applications in machine learning. *IEEE J. Sel. Areas Inf. Theory* **2020**, *1*, 19–38. [CrossRef]
23. Zaidi, A.; Estella-Aguerri, I.; Shamai (Shitz), S. On the Information Bottleneck Problems: Models, Connections, Applications and Information Theoretic Views. *Entropy* **2020**, *22*, 151. [CrossRef]
24. Lewandowsky, J.; Bauch, G.; Stark, M. Information Bottleneck Signal Processing and Learning to Maximize Relevant Information for Communication Receivers. *Entropy* **2022**, *24*, 972. [CrossRef]
25. Ben-David, S.; Blitzer, J.; Crammer, K.; Kulesza, A.; Pereira, F.; Vaughan, J.W. A theory of learning from different domains. *Mach. Learn.* **2010**, *79*, 151–175. [CrossRef]
26. Ben-David, S.; Blitzer, J.; Crammer, K.; Pereira, F. Analysis of representations for domain adaptation. In Proceedings of the International Conference on Neural Information Processing Systems, Hong Kong, China, 3–6 October 2006.
27. Pan, S.J.; Qiang, Y. A Survey on Transfer Learning. *IEEE Trans. Knowl. Data Eng.* **2010**, *22*, 1345–1359. [CrossRef]
28. Zhao, H.; Combes, R.; Zhang, K.; Gordon, G.J. On Learning Invariant Representation for Domain Adaptation. *arXiv* **2019**, arXiv:1901.09453.
29. Russo, D.; Zou, J. How much does your data exploration overfit? controlling bias via information usage. *IEEE Trans. Inf. Theory* **2019**, *66*, 302–323. [CrossRef]
30. Xu, A.; Raginsky, M. Information-theoretic analysis of generalization capability of learning algorithms. *Adv. Neural Inf. Process. Syst.* **2017**, *30*, 2521–2530.
31. Sefidgaran, M.; Gohari, A.; Richard, G.; Simsekli, U. Rate-distortion theoretic generalization bounds for stochastic learning algorithms. In Proceedings of the Conference on Learning Theory, London, UK, 2–5 July 2022; pp. 4416–4463.
32. Sefidgaran, M.; Chor, R.; Zaidi, A. Rate-Distortion Theoretic Bounds on Generalization Error for Distributed Learning. *arXiv* **2022**, arXiv:2206.02604.

33. Wang, J.; Chen, J.; Lin, J.; Sigal, L.; Silva, C.W.D. Discriminative Feature Alignment: Improving Transferability of Unsupervised Domain Adaptation by Gaussian-guided Latent Alignment. *Pattern Recognit.* **2021**, *116*. Available online: https://github.com/JingWang18/Discriminative-Feature-Alignment/tree/master/Digit_Classification/DFAMCD (accessed on 20 July 2022). [CrossRef]
34. Lecun, Y.; Bottou, L.; Bengio, Y.; Haffner, P. Gradient-Based Learning Applied to Document Recognition. *Proc. IEEE* **1998**, *86*, 2278–2324. [CrossRef]
35. Hull, J. Database for handwritten text recognition research. *IEEE Trans. Pattern Anal. Mach. Intell.* **1994**, *16*, 550–554. [CrossRef]